\documentclass[journal]{IEEEtran}

\usepackage{graphicx}
  \usepackage{here}
  \usepackage{array}
  \usepackage{color,amsmath,here}
  \usepackage{amssymb}
  \usepackage{enumerate}
  \usepackage{ascmac}
  \usepackage{tascmac}
  \usepackage{fancybox}
  \usepackage{float}
  \usepackage{amsmath}
  \usepackage{mathrsfs}
  \usepackage{xhfill}

\newtheorem{proposition}{Proposition}

\newtheorem{theorem}{Theorem}

\newtheorem{lemma}{Lemma}

\newtheorem{assumption}{Assumption}

\newtheorem{proof}{Proof}

\definecolor{mygreen}{rgb}{0, 0.7, 0}
\definecolor{myyellow}{rgb}{0.7, 0.7, 0}
\definecolor{mypurple}{rgb}{0.42, 0, 0.84}

\newcommand{\req}[1]{(\ref{#1})}

\newcommand{\dg}{\dagger}
\newcommand{\im}{{\rm im}}
\newcommand{\vc}{{\rm vec}}
\newcommand{\ol}[1]{\overline{#1}}

\ifCLASSINFOpdf
\else
\fi

\begin{document}
\title{Fast Online Reinforcement Learning Control using State-Space Dimensionality Reduction}
\author{Tomonori~Sadamoto$^1$,~\IEEEmembership{Member,~IEEE,}
        Aranya~Chakrabortty$^2$,~\IEEEmembership{Senior Member,~IEEE,}
        and~Jun-ichi~Imura$^3$,~\IEEEmembership{Senior Member,~IEEE}
	\thanks{
$^1$ Department of Mechanical and Intelligent Systems Engineering, 
Graduate School of Informatics and Engineering, The University of Electro-Communications; 1-5-1, Chofugaoka, Chofu, Tokyo 182-8585, Japan. 
Email: sadamoto@uec.ac.jp

$^2$ Electrical \& Computer Engineering, North Carolina State University; Raleigh, North Carolina, USA, 27695. 
Email: achakra2@ncsu.edu

$^3$ Department of Systems and Control Engineering, 
Graduate School of Engineering, Tokyo Institute of Technology; 2-12-1, Oh-Okayama, Meguro, Tokyo, Japan. 
Email: imura@sc.e.titech.ac.jp
}
}

\markboth{}%
{Shell \MakeLowercase{\textit{et al.}}: Bare Demo of IEEEtran.cls for Journals}

\maketitle

 \begin{abstract}
In this paper, we propose a fast reinforcement learning (RL) control algorithm that enables online control of large-scale
networked dynamic systems. RL is an effective way of designing model-free linear quadratic regulator (LQR) controllers for linear time-invariant
(LTI) networks with unknown state-space models. However,
when the network size is large, conventional RL can result
in unacceptably long learning times. The proposed approach is to construct a compressed state vector by projecting
  the measured state through a projective matrix. This matrix is
constructed from online measurements of the states in a way that it captures the dominant controllable subspace of the open-loop network model. Next, a RL-controller is learned using the reduced-dimensional state instead of the original state such that the resultant cost is close to the optimal LQR cost. Numerical benefits as well as the cyber-physical implementation benefits of the  approach are verified using illustrative examples including an example of wide-area control of the IEEE 68-bus benchmark power system. 
 \end{abstract}


\IEEEpeerreviewmaketitle

\section{Introduction}\label{sec:intro}

Learning theory has been pursued in the domain of control systems since the early 1970's, mostly through system identification and adaptive control \cite{narendra1990identification}-\cite{werbos1989neural}. 
Starting from the 90's control theorists started taking an active interest in relating learning with optimal control \cite{vidyasagar2006learning}. In particular, reinforcement learning (RL), which was originally developed in the artificial intelligence community, was related to optimal control as they both seek to find policies that minimize cost functions for a given set of tasks \cite{sutton2011reinforcement}. In recent years, RL has been refurbished with renewed enthusiasm in the context of linear quadratic regulator (LQR) designs through several papers such as
\cite{lewis2009reinforcement}-\cite{mehta2009q},
and surveys such as \cite{khargonekar2018advancing,recht2018tour}. 

The fundamental idea behind RL-based LQR is to iteratively learn the optimal state-feedback control gain matrix directly from online measurements of the states and inputs without knowing the system model.
Among existing RL techniques such as value iteration \cite{wei2015value}-\cite{bertsekas1995dynamic}, a technique known as {\it policy iteration} has fast convergence \cite{sutton2011reinforcement} when a given policy structure matches the optimal solution. In policy iteration, the system is initially excited with exploratory noise, followed by gathering sufficient number of state and input samples for updating the parameters of the policy. For LQR this parameter update is performed by least squares.  This estimate is thereafter iterated upon till the estimate converges to produce the optimal control gain, amounting to solving the algebraic Riccati equation using state and input data. However, even for a reasonably fast method such as policy iteration, curse of dimensionality continues to be an ongoing debate. One main reason behind this is that the least squares step requires at least $n(n+1)/2$ number of data samples for a unique solution where $n$ is the order of the system. As a result, the computational cost of policy iteration for every iterate is $\mathcal O(n^6)$.  This can pose a serious challenge for real-time decision-making and control in networks with large values of $n$.  


A few approaches have been proposed to overcome this curse. For example, in \cite{jiang2012robust,jiang2012robust2} the authors have proposed a method to learn decentralized RL-controllers by regarding the interference between various subsystems in a network as an uncertainty. This method, however, is mostly applicable for weakly-connected networks.
In the artificial intelligence community, on the other hand, a notion called {\it state aggregation} has been proposed to overcome the computational bottleneck of RL in controlling Markov decision processes (MDP). The idea here is that instead of the original state-space the decision-maker finds decision vectors in an abstract  state-space much faster by treating groups of states as a unit, ignoring irrelevant state information. A number of abstractions have been proposed, for example see \cite{tsitsiklis1996feature,singh1995reinforcement,lin2010evolutionary,abel2017near}, with a brief survey in \cite{li2006towards}. However, these abstraction methods are not based on any control-theoretic property of the network, nor do they answer quantitative questions such as ``what system-theoretic information is lost when an abstraction is applied?'', or ``how do the dynamics of the network model decide the level of abstraction'', etc.

In this paper we propose a new approach for fast design of RL controllers using the concept of dimensionality reduction. The basic idea is to exploit low-rank property of the controllability gramian of the system, project the measured states into a corresponding lower-dimensional space that captures the dominant eigenvectors of the gramian, and learn a LQR controller in this lower-dimensional space.
Low-rank property of the controllability gramian in this case means that
following a disturbance, one would need to control only the {\it dominantly controllable} behavior of the network
states, and still be able to steer the network to its desired mission.
Many practical systems that have lesser number of control inputs than states exhibit such low-rank controllability property. Common examples include consensus networks \cite{sadamoto2016average}, heat diffusion networks \cite{sadamoto2016average,sadamoto2014nonlinear}, electric power systems \cite{chow2013power} and transportation networks \cite{niedbalski2008model,hoogendoorn2001state}. Especially in power system applications, several recent results have proposed the use of low-rank controllability for designing broadcast type controls, two leading examples being load-frequency control \cite{shayeghi2009load} and power oscillation damping \cite{xue2019control}.
Based on these observations, our proposed approach is to construct a matrix from online measurements of states and inputs that projects these measurements to a low-dimensional data space, capturing the dominant traits of the network dynamics. The matrix construction is done using singular value decomposition of the measurements. A RL controller is then designed based on this compressed data. Because of its lower dimensionality, the computational complexity for learning is drastically reduced. Stability and optimality of the control performance are theoretically evaluated using robust control theory by treating the dimensionality-reduction error as an uncertainty. The effectiveness of the proposed learning method is illustrated by numerical simulations. 

A dimensionality-reduction based RL controller was recently proposed in \cite{sayakcdc} under the assumption that the network model exhibits a singular perturbation structure. This method is a special case of our approach when the projection matrix is known and is chosen to extract the slow-time-scale dynamics.  Model-based control designs using dimensionality reduction have been proposed in earlier papers such as \cite{kokotovic1999singular} as well as in more recent papers such as \cite{xue2016optimal}. However, their extension to a completely model-free execution such as ours has not been reported yet. Our method is also superior to control designs that are based on system identification as in \cite{watter2015embed} as identification-based methods are indirect and so one cannot guarantee optimality even if full-dimensional system models are constructed. In contrast, our approach guarantees \{suboptimality, optimality\} when the dimensionality approximation error is \{small, zero\}. 

The rest of the paper is organized as follows. Section \ref{sec:prob} formulates the problem of fast RL control design, followed by Section \ref{sec:propose} which presents an algorithm for this design assuming that the dimensionality reduction is exact. Theoretical results on closed-loop stability and performance degradation depending on the reduction error are derived in Section \ref{sec:relax}.  An extension of the proposed algorithm to networks with semi-stable dynamics is presented in Section V, followed by numerical simulations in Section \ref{sec:sim}.  Section \ref{sec:con} concludes the paper.

{\bf Notation:} We denote the set of real numbers as $\mathbb R$,
the pseudo-inverse of a full-row rank matrix $P$ as $P^{\dg}$,
the range space spanned by the column vectors of a matrix $P$ as $\im P$,
the null space spanned by those as $\ker P$, and
the $n$-dimensional identity matrix as $I_n$. The subscript $n$ is omitted if obvious.
We denote the $i$-th column vector of $I_m$ as $e_i^{m}$, and the $n$-dimensinoal row vector whose entries are all one as ${\bf 1}_n$.
We denote a positive definite (semi-definite) matrix $A$ as $A \succ 0$ ($A \succeq 0$). 
Given $P \in \mathbb R^{\hat{n}\times n}$ where $\hat{n} \leq n$, the matrix $\ol{P}$ is defined such that $P^{\dg}P + \ol{P}^{\dg}\ol{P} = I$. When $\hat{n} = n$, $\ol{P}$ is empty.
The operator $\otimes$ denotes the Kronecker product.
For a matrix $P := [p_1,\ldots, p_n]$,  $\vc (P) := [p_1^{\sf T}, \ldots, p_n^{\sf T}]^{\sf T}$.
For a given Hurwitz $A \in \mathbb R^{n \times n}$ and $B \in \mathbb R^{n \times m}$, the controllability gramian is defined as $\Phi(A,B) := \int_0^{\infty} e^{At}BB^{\sf T}e^{A^{\sf T}t}dt$. When $\Phi(A,B)$ is full rank, the dynamical system $\dot{x} = Ax + Bu$ is said to be controllable.
The Dirac-delta function is denoted as $\delta(t)$.
The $\mathcal L_2$-norm of a square integrable function $v(t) \in \mathbb{R}^n$ is defined by $\|v(t)\|_{\mathcal L_2} := \left(\int_{0}^{\infty} v^{\sf T}(t)v(t) dt\right)^{\frac{1}{2}}$. 
We denote the set of integrable function $v(t)$ satisfying $\|v(t)\|_{\mathcal L_2} \leq 1$ as $\mathcal L_2^{\rm nor}$. 
The $\mathcal{H}_{\infty}$-norm of a stable proper transfer matrix $G(s)$ is defined by $\|G(s)\displaystyle \|_{\mathcal{H}_{\infty}}:=\textstyle \sup_{\omega\in \mathbb{R}}\|G(j\omega)\|$ where $\|\cdot\|$ denotes the induced $2$-norm, and the $\mathcal{H}_{2}$-norm of a stable strictly proper transfer matrix $G(s)$ is defined by $\|G(s)\|_{\mathcal{H}_{2}}:= \textstyle \left(\frac{1}{2\pi}\int_{-\infty}^{\infty}{\rm tr}(G(j\omega)G^{{\sf T}}(-j\omega))d\omega\right)^{\frac{1}{2}}$. 

\section{Problem Setup}\label{sec:prob}
\subsection{Brief Review of Off-Policy Iteration}\label{sec:probset}
Consider a linear networked dynamical system consisting of $L$ subsystems. For $l \in \{1,\ldots, L\}$, the $l$-th subsystem dynamics is described as
\begin{equation}\label{subsys}
 \Sigma_{l}: ~\dot{x}_{[l]} = A_{ll}x_{[l]} + \sum_{j \in \mathcal N_{l}} A_{lj} x_{[j]} + B_{l}u_{[l]},\quad
  x_{[l]}(0) = x_{[l]0}
\end{equation}
where $x_{[l]}\in\mathbb R^{n_{l}}$ is a state, $u_{[l]} \in \mathbb R^{m_{l}}$ is a control input, $\mathcal N_{l}$ is the index set of subsystems connecting to the $l$-th subsystem, and $x_{[l]0}$ is an initial state of the subsystem. Using the notation
\begin{equation}
 \begin{array}{ll}
 A := \left[
\begin{array}{ccc}
 A_{11}&\cdots&A_{1 L} \\
 \vdots &\ddots& \vdots \\
 A_{L 1}&\cdots&A_{L L} \\
\end{array}
      \right], &
B := \left[
\begin{array}{ccc}
 B_{1}& & \\
 & \ddots & \\
 && B_{L} 
\end{array}
\right]
\\
  n := \sum_{l = 1}^{L} n_{l}, &
  m := \sum_{l = 1}^{L} m_{l},  \vspace{1mm}\\
  x := [x_{[1]}^{\sf T},\ldots, x_{[L]}^{\sf T}]^{\sf T}, &
 u := [u_{[1]}^{\sf T},\ldots, u_{[L]}^{\sf T}]^{\sf T}, \\
 \end{array}
\end{equation}
the interconnected network model can be written as
\begin{equation}\label{sys}
 \Sigma: ~\dot{x} = Ax + Bu, \quad x(0) = x_0. 
\end{equation}
We impose the following three assumptions on \req{sys}.
\vspace{2mm}
\begin{assumption}
 The matrices $A$ and $B$ are unknown.
\end{assumption}
\vspace{2mm}
\begin{assumption}
 The state $x$ is measurable.
\end{assumption}
\vspace{2mm}
\begin{assumption}
 The open-loop system in \req{sys} is  stable, i.e., $A$ is Hurwitz.
\end{assumption}
\vspace{2mm}
Assumption 1 implies that although we know that the system of our interest is LTI we do not know its model.
Even though many control systems in practice are nonlinear, their behavior can often be captured quite accurately by a linearized model around an operating point. In that case, this assumption implies that this linearized model is unknown, either because the original nonlinear model is unknown or because of other practical uncertainties. 
%
Assumptions 1-2 imply that both $n$ and $m$ are known.
Assumption 3 is usually satisfied in many real-world networks where the underlying physical laws (Newton's law, Faraday's law, Kirchoff's law, etc.) ensure that the network dynamics are stable even if the network parameters are unknown. 
Our goal is to design a state-feedback controller
\begin{equation}\label{con}
 u = -Fx
\end{equation}
such that the cost function
\begin{equation}\label{Jx0}
 J := \int_{0}^{\infty} x^{\sf T}(t)Qx(t) + u^{\sf T}(t)Ru(t) dt
\end{equation}
where $Q \succeq 0$ and $R \succ 0$ are chosen matrices, is minimized. If $A$ and $B$ were known  one can find the optimal controller $F$ in \req{con} by solving an algebraic Riccati equation \cite{lewis2012optimal}. However, since both of these matrices are completely unknown, the standard LQR approach no longer applies. Instead a reinforcement learning (RL) based approach needs to be used, as shown in recent papers such as \cite{jiang2012robust,vrabie2009adaptive}. 

Several variants of RL exist in the literature.
One popular method is known as policy iteration, which is further categorized into two approaches: on-policy and off-policy iteration \cite{sutton2011reinforcement,jiang2012robust}. Both on- and off-policy iteration algorithms are equivalent to a data-driven implementation of the {\it Kleinman's Algorithm}, which is an iterative scheme to solve the algebraic Riccati equation associated with \req{Jx0} \cite{kleinman1968iterative}. At the $k$-th step of the on-policy iteration, under the assumption that a stabilizing control gain $F_k$ is obtained, one updates the gain based on the data collected after applying the control law. This means that one has to recollect the data every time one updates the control law, resulting in a long learning time. Off-policy iteration, on the other hand, enables one-shot learning, i.e. one can find the optimal control gain by using only one set of data obtained before actuating the control signal. Because of its faster speed, we will be using off-policy iteration for developing our proposed RL controller. We  next recapitulate the basic steps of the off-policy method.  

The off-policy iteration consists of two stages, namely - {\it Online Data Collection} and {\it Policy Improvement}. In the first stage, one excites the system \req{sys} with an exploration noise, say denoted as $u(t)=w(t)$ for $0 \leq t \leq 1$, $u(t) = 0$ for $t > 1$, and collects the time-series measurements of $x(t)$ driven by this noise and the initial state disturbance $x_0$. In the second stage, one estimates the optimal control gain iteratively using the measured data. An overview of this estimation is as follows. Note that the open-loop system \req{sys} can be rewritten as
 \begin{equation}\label{PI_con_OFF_01}
  \dot{x} = A_k x + Bu', \quad A_k := A + BF_k, \quad u' := u - F_k x
 \end{equation}
 where $F_k$ is a given stabilizing control gain, i.e., $A + BF_k$ is Hurwitz. This fictitious representation allows us to interpret the sequence $x(t)$ following the open-loop system \req{sys} as the data of the closed-loop system \req{PI_con_OFF_01} with an external signal $u'$. Given $\{t_j\}_{j=0}^{N}$, an incremental Lyapunov function candidate can be written as 
\begin{align}     
    &\scalebox{0.92}{$\displaystyle x^{\sf T}(t_{j+1})P_kx(t_{j+1}) - x^{\sf T}(t_j)P_kx(t_j) = \int_{t_j}^{t_{j+1}} \frac{d}{d\tau}x^{\sf T}(\tau)P_kx(\tau)d\tau  $} \nonumber \\
     =& \scalebox{0.92}{$\displaystyle -\int_{t_j}^{t_{j+1}} x^{\sf T}(\tau) Q_k x(\tau) d\tau  - 2 \int_{t_j}^{t_{j+1}}u'^{\sf T}(\tau)RF_{k+1}x(\tau) d\tau $} \label{OFF_PI_derive02}
\end{align}
where $Q_k := Q + F_k^{\sf T}RF_{k}$. Note that \req{OFF_PI_derive02} follows from the Kleinman's Algorithm. Because $x(t)$ in \req{OFF_PI_derive02} follows from the open-loop system driven by the exploration noise $u(t)$, one only needs to collect $x(t)$ once for computing $F_k$. Once $F_k$ converges, the final controller is implemented. The complete off-policy iteration algorithm is summarized in Algorithm 1. 


\noindent\rule{\columnwidth}{1pt}
{\bf Algorithm 1:} Off-policy iteration \cite{jiang2012robust}\vspace{-1.5mm}

\noindent\rule{\columnwidth}{0.4pt}
\noindent {\bf Initialization:}

\noindent \hspace{2mm} Given $N$, $j\leftarrow 0$, $k \leftarrow 0$,  ${F}_0 = 0$, $\kappa \geq 0$, fix a set $\{t_j\}_{j\in \{0,\ldots,N\}}$ such that $t_{N} > \ldots > t_0 \geq 0$.  \vspace{0.5mm}

\noindent {\bf Online Data Collection:}

\noindent \hspace{1.2mm} For $j = 0, \ldots, N-1$, do:
\begin{itemize}
 \item[1.] Measure $x(t)$ for $t \in [t_j, t_{j+1}]$ under a given $u(t)$.
 \item[2.] Numerically compute
 \begin{eqnarray}
  \hspace{-8mm}
 \phi_j &\hspace{-2mm}:=&\hspace{-2mm} \left(x(t_{j+1}) \otimes x(t_{j+1}) - x(t_{j}) \otimes x(t_{j})\right)^{\sf T}  \in \mathbb R^{1 \times n^2} \label{phi} \\
  \hspace{-8mm} \rho_j &\hspace{-2mm}:=&\hspace{-2mm} \int_{t_j}^{t_{j+1}} \left(x(t) \otimes u(t)\right)^{\sf T} dt \in \mathbb R^{1 \times nm} \label{rho} \\
  \hspace{-8mm} \sigma_j &\hspace{-2mm}:=&\hspace{-2mm} \int_{t_j}^{t_{j+1}} \left(x(t) \otimes x(t)\right)^{\sf T} dt  \in \mathbb R^{1 \times n^2} \label{sigma}
 \end{eqnarray}
\end{itemize}
\vspace{1mm}

\noindent {\bf Policy Improvement:}
\begin{itemize}
 \item[3.] Find ${W}_k$ and ${F}_{k+1}$ satisfying
\begin{equation}\label{sol}
 \Theta_k\left[
  \begin{array}{c}
   \vc({W}_k)\\    \vc({F}_{k+1})
  \end{array}
  \right] = z_k
\end{equation}
with
\begin{eqnarray}
\hspace{-8mm}\scalebox{0.92}{$\displaystyle \Theta_k$} &\hspace{-2mm}:=&\hspace{-2mm}  \scalebox{0.92}{$\displaystyle \left[\phi~ -2\rho (I \otimes R)-2\sigma(I \otimes {F}_k^{\sf T}R)\right] \in \mathbb R^{N \times (2n^2+nm)} $}\label{Theta} \\
\hspace{-8mm} z_k &\hspace{-2mm}:=&\hspace{-2mm} -\sigma \vc({Q} + {F}_k^{\sf T}R{F}_k)\in \mathbb R^N 
\end{eqnarray}
where $\phi \in \mathbb R^{N \times n^2}$, $\rho \in \mathbb R^{N \times nm}$, and $\sigma \in \mathbb R^{N \times n^2}$ are the stacked versions of $\phi_j$, $\rho_j$, and $\sigma_j$, respectively.
\item[4.] Exit if $\|{F}_{k+1} - F_{k}\| \leq \kappa$, otherwise let $k\leftarrow k+1$ and return to 3.
\end{itemize}
\noindent\rule{\columnwidth}{1pt}
\vspace{2mm}

Note that \req{sol} is the least-squares representation of \req{OFF_PI_derive02}. The following theorem from \cite{jiang2012robust} shows that the off-policy iteration in Algorithm 1 produces an optimal controller. \vspace{1mm}
\begin{theorem}\cite{jiang2012robust}
 Consider $\Sigma$ in \req{sys} and $J$ in \req{Jx0}. Assume
 \begin{equation}\label{rankcond}
  {\rm rank}[\rho ~ \sigma] = \frac{{n}({n}+1)}{2} + m{n}
 \end{equation}
 where $\rho$ and $\sigma$ are the stacked vector representations of $\rho_j$ in \req{rho} and $\sigma_j$ in \req{sigma}. Then, the following two statements hold:
 \begin{enumerate}
  \item[i)] $A-BF_k$ is Hurwitz for any integer $k>0$, where $F_k$ follows from the solution of \req{sol}.
  \item[ii)] Define $F_{\infty} := \lim_{k \rightarrow \infty}F_k$. The control $u = -F_{\infty}x$ minimizes $J$.
 \end{enumerate}
\end{theorem}
\vspace{1mm}


Since Algorithm 1 is equivalent to Kleinman's Algorithm \cite{kleinman1968iterative}, it can be easily derived that the iterates $W_k$ are quadratically convergent. 
The main limitation of Algorithm 1 is that it is not easily scalable to high-dimensional networks with a large value of $n$. Step 3 of the algorithm is the main computational bottleneck. To solve \req{sol}, one needs to compute the pseudo-inverse of $\Theta_k$ in \req{Theta}. The computational cost for that based on singular value decomposition is  $\mathcal O(\min\{N^2n^4, n^2N^4\})$ \cite{holmes2007fast}, where $N$ is the number of collected data samples. 
To satisfy \req{rankcond}, one must also have $N \geq \frac{{n}({n}+1)}{2} + m{n}$, which makes the overall cost of the algorithm $\mathcal O(n^6)$. Moreover, the pseudo-inverse must be computed at every iteration of the policy improvement. Thus, the {\it learning time}, defined as the time needed for running the Policy Improvement of Algorithm 1, may be unacceptably large, thereby limiting its application for real-time control.

\subsection{Purpose and Approach of This Paper}\label{sec:pur}

Motivated by this problem, in this paper we propose an alternative approach for designing the model-free controller \req{con} such that it makes the cost function $J$ in \req{Jx0} as small as possible while reducing the learning time to significantly lesser than $\mathcal O(n^6)$.
Our fundamental approach is to utilize the low-rank property of the controllability gramian of \req{sys}. In many large-scale network systems, disturbances and control inputs tend to excite only a part or a combination of the state variables \cite{antoulas2005approximation}, implying  that the collected data $x(t)$  associated with a low-dimensional controllable subspace is sufficient for learning. As long as the data is compressed appropriately from the viewpoint of controllability, learning based on the low-dimensional data can be expected to work well. Our goal is to theoretically establish this intuition while at the same time developing an algorithm to learn $F$ using low-dimensional data. The proposed approach is briefly summarized as follows:
\begin{enumerate}
 \item First, we construct a reduced-dimensional compressed state vector $\xi$ as
       \begin{equation}\label{defXI_first}
	\xi := Px,
       \end{equation}
       by projecting the measured state trajectory $x(t)$ through a full row-rank matrix $P \in \mathbb R^{\hat{n} \times n}$ where $\hat{n} \ll n$. The matrix $P$ will capture the level of redundancy in the controllability of the network model \eqref{sys} that allows for dimensionality reduction. Note that the system model is not known, so $P$ will be constructed solely from the measurements of $x(t)$.   
       
 \item Next, a controller \req{con} is learned using the reduced-dimensional state $\xi(t)$  instead of the original state $x(t)$ such that the performance cost of the resulting closed-loop system is as close to the optimal cost $J$ in \req{Jx0} as possible.
\end{enumerate}
In the following sections, we show how these steps are achieved. 


 \section{Proposed Algorithm} \label{sec:propose}
 \subsection{A Sufficient Condition for Lossless Compression}
We first consider an ideal situation when the data compression is lossless, i.e. $x(t)$ can be recovered from the compressed data $\xi(t)$. In that case, one can expect that the learning based on the lossless data $\xi(t)$ will provide an optimal controller minimizing $J$ in \req{Jx0}.  This is indeed true, as shown next.
A sufficient condition for the lossless compression is to reduce the redundant states of $x$ in terms of controllability. This is summarized as the following lemma.

\vspace{2mm}
\begin{lemma}
 Consider $\Sigma$ in \req{sys}. Let $P$ satisfy
  \begin{equation}\label{Pcon}
   {\rm im} P^{\sf T} = {\rm im}~\mathcal R(A, [B~ x_0])
  \end{equation}
where $\mathcal R(A,b) := [b, Ab, \ldots, A^{n-1}b]$.
 Then, $\xi$ in \req{defXI_first} satisfies
 \begin{equation}\label{lowsys}
  \hat{\Sigma}:~ \dot{\xi} := PAP^{\dg}\xi + PBu, \quad \xi(0)=Px_0,
 \end{equation}
 where $PAP^{\dg}$ is Hurwitz. 
 Furthermore,
 \begin{equation}\label{xxi}
  x(t) \equiv P^{\dg}\xi(t), \quad \forall t
 \end{equation}
 holds for any $u(t)$ and $x_0$.
\end{lemma}
\vspace{2mm}
 \begin{proof} See  \cite{kalman1962canonical}. 
 \end{proof}
\vspace{2mm}

This lemma implies that if $P$ satisfies \req{Pcon}, the $\hat{n}$-dimensional state $\xi$ exactly captures the behavior of the $n$-dimensional state $x$.  From \req{xxi}, it follows that the cost function with respect to the compressed state $\xi$
\begin{equation}\label{Jhat}
 \hat{J} := \int_0^{\infty} \xi^{\sf T}(t)\hat{Q}\xi(t) + u^{\sf T}(t)Ru(t) dt,\quad \hat{Q} := (P^{\dg})^{\sf T}QP^{\dg}
\end{equation}
is identical to $J$ in \req{Jx0}. In other words, the learning for $n$-dimensional system $\Sigma$ in \req{sys} with a cost function $J$ by using $\xi$ is equivalent to the learning for the reduced-order system \req{lowsys} with a cost function $\hat{J}$ in \req{Jhat}. This can be done by employing the off-policy iteration method of Algorithm 1 if a matrix $P$ satisfying \req{Pcon} is found. 

\subsection{Construction of $P$ from Data}
We next show an approach by which one can construct  $P$ satisfying \req{Pcon} by using the measured data $x(t)$ instead of the system model $A$ and $B$. We first introduce the following lemma.
\vspace{2mm}
 \begin{lemma}
  Consider $\Sigma$ in \req{sys}. If $P$ satisfies \req{Pcon}, the condition
   \begin{equation}\label{exact_P}
    (I - P^{\dg}P)x(t) \equiv 0, \quad \forall t
   \end{equation}
  holds for any $u$ and $x_0$. In addition, without loss of generality, we suppose that $\hat{n}$ is the minimal dimension for satisfying \req{exact_P}. Then, \req{exact_P} is equivalent to \req{Pcon}. 
 \end{lemma}
\vspace{2mm}
   \begin{proof}
    First, we show the sufficiency. If \req{Pcon} holds, by multiplying $\ol{P}^{\dg}\ol{P}$ to \req{xxi} from the left, we have $\ol{P}^{\dg}\ol{P}x(t) \equiv 0$ for any $u$, $x_0$, $t$. This is equivalent to \req{exact_P}. 
   Next, we show \req{Pcon} if \req{exact_P} holds.
   To this end, we first show
  \begin{equation}\label{firstshoe}
   P^{\dg}P\mathcal R(A, [B~x_0]) = \mathcal R(A, [B~x_0]). 
  \end{equation}
    Note that \req{exact_P} is assumed to be satisfied when $u=0$. Thus, we have
  \begin{equation}\label{Px0_in}
   P^{\dg}P e^{At}x_0 =  e^{At}x_0, \quad \forall t.
  \end{equation}
   Substituting $t=0$ in \req{Px0_in}, we have $P^{\dg}Px_0 = x_0$. Next, taking one time derivative of \req{Px0_in} and substituting $t=0$, we have $P^{\dg}PAx_0 = Ax_0$. Repeating this procedure up to the $(n-1)$-th order time-derivative, we have $P^{\dg}PA^{i}x_0 = A^ix_0$ for $i \in \{0,\ldots, n-1\}$. Next, consider an impulse input applied to the $i$-th control port, described as  
   \begin{equation}\label{uimp}
  u(t) = e_{i}^{m}\delta(t). 
   \end{equation}
    Note that $x(t)$ under this impulse input is $ x(t) = e^{At}(x_0 + B_i)$. Thus, \req{exact_P} can be written as 
  \[
  P^{\dg}P e^{At}(x_0+B_i)  =  e^{At}(x_0+B_i), \quad \forall t,
  \] 
    where $B_i$ denotes the $i$-th column of $B$. Substituting $t=0$, we have $P^{\dg}PB_i = B_i$, where the relation $P^{\dg}Px_0 = x_0$ is used. Repeating the same procedure for all $i \in \{1,\ldots, m\}$ and up to the $(n-1)$-th order time-derivative, the relation \req{firstshoe} follows.
    Clearly, \req{firstshoe} is equivalent to ${\rm im} P^{\sf T} \supseteq {\rm im} \mathcal R(A, [B~x_0])$. A minimal-dimensional $P$ satisfying this relation yields ${\rm im} P^{\sf T} = {\rm im} \mathcal R(A, [B~x_0])$, which is \req{Pcon}. This completes the proof. \hspace{\fill}$\square$
   \end{proof}
\vspace{2mm}

Lemma 2 implies that $P$ satisfying \req{exact_P} spans a reachable subspace from $u$ and $x_0$.
However, finding a $P$ which satisfies \req{exact_P} exactly for any arbitrary input signal $u$ can be difficult. Therefore, we consider solving the following online minimization problem:  
\begin{equation}\label{minpro}
 \min_{P \in \mathbb R^{\hat{n} \times n}} \sum_{j=0}^{N-1} \|(I-P^{\dg}P) x(t_j)\|^2, 
\end{equation}
where $\hat{n} \leq n$ is a given design parameter, and $x(t_j)$ is the sampled state driven by initial state disturbance $x_0$ and an exploratory noise $u(t)$.
The $x(t)$ measurements that are collected during the {\it Online Data Collection} step of Algorithm 1 can be used in \req{minpro} for finding $P$.
One can see that the solution of \req{minpro} approximately satisfies \req{exact_P} if $N$ is sufficiently large. 
This minimization problem can be solved by using the singular value decomposition of
\begin{equation}\label{defXX}
X := [x(t_0), \ldots, x(t_{N-1})] \in \mathbb R^{n \times N},
\end{equation}
with $P^{\dg}$ given by the first $\hat{n}$ columns of the left-singular matrix.

Another option can be to construct $P$ offline, i.e., before the disturbance hits the network. This can further save online computation time. Offline construction of $P$, for example, can be done by using the {\it empirical controllability gramian} \cite{lall1999empirical}. For $i \in \{1,\ldots, m\}$, let $x_i(t)$ be the response of the system $\Sigma$ in \req{sys} with the impulse input given by \req{uimp}. The empirical controllability gramian is defined as
\begin{equation}\label{Phiemp}
 \Phi_{\rm emp}(t) := \sum_{i=1}^{m} \int_0^{t}
  \left(x_i(\tau) - \bar{x}_i(t)\right)  \left(x_i(\tau) - \bar{x}_i(t)\right)^{\sf T} d\tau
\end{equation}
with $\bar{x}_i(t) := \frac{1}{t}\int_0^{t}x_i(\tau)d\tau$. It can be clearly seen that
\begin{equation}
 \lim_{t\rightarrow \infty}\Phi_{\rm emp}(t) = \Phi(A,B). 
\end{equation}
This implies that we can construct the controllability gramian by using $m$ sets of $\{x(t), u(t)\}$ without knowing the system model.  
  Although we do not know what disturbance is injected, it may be possible to know a possible subspace $\mathcal X \subseteq \mathbb R^n$ in which $x_0$ lies. In this situation, similarly as the above, we can construct another empirical controllability gramian associated with the disturbance. Let $d_1, \cdots, d_r$ be the basis of the distribution $\mathcal X$, i.e., ${\rm im} [d_1, \ldots, d_r] = \mathcal X$. Let $x_i(t)$  be the response of $\Sigma$ with the disturbance of $x(0) = d_i$ for $i \in \{1,\ldots, r\}$. Denote the empirical controllability gramian \req{Phiemp} by $\Phi_{\rm emp}^{\rm x}$. Similarly, denote the gramian \req{Phiemp} associated with the input by $\Phi_{\rm emp}^{\rm u}$.  Then, by choosing eigenvectors of $\Phi_{\rm emp} := \Phi_{\rm emp}^{\rm u} + \Phi_{\rm emp}^{\rm x}$ corresponding to the non-zero eigenvalues, we can construct $P$ satisfying \req{firstshoe} for any $x_0 \in \mathcal X$.  

 
\subsection{Preconditioned Off-Policy Iteration}\label{sec:redundant}
Based on the idea of constructing the projection matrix $P$ from $x(t)$, and compressing $x$ to $\xi$ as described above, we are now ready to present our proposed RL control method. We refer to this method as  {\it preconditioned} off-policy iteration. Unlike the regular off-policy iteration this method consists of three stages, namely - {\it Online Data Collection}, {\it Preconditioning}, and {\it Policy Improvement}. The first stage is identical to that in Algorithm 1.
In the second stage,  we define $\hat{\phi}$, $\hat{\rho}$, $\hat{\sigma}$ by replacing $x$ in \req{phi}-\req{sigma} with $\xi$. The revised matrices are computed as
\begin{eqnarray}
 \hat{\phi}  &\hspace{-2mm} =&\hspace{-2mm} \phi (P^{\sf T} \otimes P^{\sf T}) \in \mathbb R^{1 \times \hat{n}^2}\label{phihat}\\
 \hat{\rho} &\hspace{-2mm} =&\hspace{-2mm} \rho (P^{\sf T} \otimes I) \in \mathbb R^{1 \times m\hat{n}} \label{rhohat}\\
 \hat{\sigma} &\hspace{-2mm} =&\hspace{-2mm} \sigma (P^{\sf T} \otimes P^{\sf T}) \in \mathbb R^{1 \times \hat{n}^2}\label{sigmahat}
\end{eqnarray}
The third stage is similar to the policy improvement stage of Algorithm 1, but the vector $x$ is replaced with $\xi$ by using the relation \req{xxi}. The pseudo-code of the proposed algorithm is shown in {\bf Algorithm 2}. 

\noindent\rule{\columnwidth}{1pt}
{\bf Algorithm 2:} Preconditioned off-policy iteration\vspace{-1.5mm}

\noindent\rule{\columnwidth}{0.4pt}
\noindent {\bf Initialization and Online Data Collection:}

Same as in Algorithm 1.

\noindent {\bf Preconditioning:}

\begin{itemize}
 \item[1.] Given $\hat{n}$, find $P$ by SVD of \req{defXX}. 
 \item[2.] Compute $\hat{\phi}$, $\hat{\rho}$, and $\hat{\sigma}$ in \req{phihat}-\req{sigmahat}.
\end{itemize}

\noindent {\bf Policy Improvement:}
\begin{itemize}
 \item[3.] Find $\hat{W}_k$ and $\hat{F}_{k+1}$ satisfying
\begin{equation}\label{sol2}
 \hat{\Theta}_k\left[
  \begin{array}{c}
   \vc(\hat{W}_k)\\    \vc(\hat{F}_{k+1})
  \end{array}
  \right] = \hat{z}_k
\end{equation}
with
\begin{eqnarray}
\hspace{-9mm}\scalebox{0.92}{$\displaystyle \Theta_k$} &\hspace{-2mm}:=&\hspace{-3mm}  \scalebox{0.92}{$\displaystyle \left[\hat{\phi}~ -2\hat{\rho} (I \otimes R)-2\hat{\sigma}(I \otimes \hat{F}_k^{\sf T}R)\right] \in \mathbb R^{N \times (2\hat{n}^2+\hat{n}m)} $}\label{Thetahat} \\
 \hspace{-9mm} \hat{z}_k &\hspace{-2mm}:=&\hspace{-3mm} -\hat{\sigma} \vc(\hat{Q} + \hat{F}_k^{\sf T}R\hat{F}_k)\in \mathbb R^N
\end{eqnarray}
where $\hat{\phi}$, $\hat{\rho}$, $\hat{\sigma}$ are defined as \req{phihat}-\req{sigmahat}, and $\hat{Q}$ is in \req{Jhat}. 
 \item[4.] Go to the next step if $\|\hat{F}_{k+1} - \hat{F}_{k}\| \leq \kappa$, otherwise, let $k \leftarrow k+1$ and return to 3.
 \item[5.] Return
	   \begin{equation}\label{defK}
	    F_{k} = \hat{F}_{k} P
	   \end{equation}
\end{itemize}
\noindent\rule{\columnwidth}{1pt}
\vspace{0mm}

Similar to Proposition 1, the convergence of Algorithm 2 and the optimality of the final controller are guaranteed as follows.

\vspace{2mm}
\begin{theorem}
Consider $\Sigma$ in \req{sys}, and a  $P$ that satisfies \req{exact_P}. Assume
  \begin{equation}\label{rankcond2}
  {\rm rank}[\hat{\rho} ~ \hat{\sigma}] = \frac{{\hat{n}}(\hat{n}+1)}{2} + m\hat{n},
 \end{equation}
 where $\hat{\rho}$ and $\hat{\sigma}$ are defined in \req{rhohat}-\req{sigmahat}.
 If Algorithm 2 is applied to $\Sigma$, then the following two statements hold:
 \begin{enumerate}
  \item[i)] $A-BF_k$ is Hurwitz for any integer $k>0$, where $F_k$ is given by \req{defK}.
  \item[ii)] Define $F_{\infty} := \lim_{k \rightarrow \infty}F_k$. The control $u = -F_{\infty}x$ minimizes $J$ in \req{Jx0}.
 \end{enumerate}
\end{theorem}

\vspace{2mm}
 \begin{proof}
  From the proof of Theorem 2.3.12 in \cite{jiang2012robust}, it follows that the control $u = -\hat{F}_k\xi$ stabilizes $\hat{\Sigma}$ in \req{lowsys}, i.e.,  $PAP^{\dg} - PB\hat{F}_k$ is Hurwitz for any integer $k>0$. Let $\ol{\xi} := \ol{P}x$.  Applying the coordinate transformation from $x$ to $[\xi^{\sf T}, \ol{\xi}^{\sf T}]^{\sf T}$, \req{sys} is written
as
    \begin{equation}\label{lowsys1}
  \left[
\begin{array}{c}
 \dot{\xi} \\
 \dot{\ol{\xi}}
\end{array}
  \right] =
  \left[
   \begin{array}{cc}
    PAP^{\dg} & PA\ol{P}^{\dg}\\
    0 & \ol{P}A\ol{P}^{\dg}
   \end{array}
\right]  \left[
\begin{array}{c}
 {\xi} \\
 {\ol{\xi}}
\end{array}
\right] +
\left[
\begin{array}{c}
 PB \\
 0
\end{array}
  \right]u,
  \end{equation}
  with $\xi(0) = Px_0$ and $\ol{\xi}(0) = 0$. Note that $\ol{P}B = 0$, $\ol{P}x_0 = 0$, and $\ol{P}AP^{\dg} = 0$ follow from \req{Pcon}. Since $A$ is Hurwitz, from the upper-triangular structure of \req{lowsys1}, $\ol{P}A\ol{P}^{\dg}$ is also Hurwitz. Furthermore, since $PAP^{\dg} - PB\hat{F}_k$ is Hurwitz, we can see that $A-BF_k$ is Hurwitz for any $k$.  The claim i) follows from this. Furthermore, the control law $u = -\hat{F}_{\infty}\xi$ minimizes \req{Jhat}  subject to the dynamics  \req{lowsys}. Finally, due to the uncontrollability of $\ol{\xi}$ as shown in \req{lowsys1} the control minimizes $J$ in \req{Jx0}. This completes the proof. \hspace{\fill}$\square$
 \end{proof}
\vspace{2mm}
 
 The computational complexity of Algorithm 2 is significantly lower than that of Algorithm 1.
 Note that the left-singular eigenvectors of $X$ are identical to those of $XX^{\sf T}$. The complexity of the preconditioning step is $\mathcal O(\max\{Nn, n^3\})$ where the first term is for computing $XX^{\sf T}$ and the second is for finding $P$ by the SVD of $XX^{\sf T}$ \cite{holmes2007fast}. In reality, the value of $N$ for capturing the dominant controllable subspace will not be large; we will show this later in our numerical simulations. Methods such as \cite{lee2014very} for computing dominant left-singular vectors can further reduce this computational cost. As mentioned in Section~\ref{sec:probset}, the computational cost for solving \req{Thetahat} is $\mathcal O(\hat{n}^6)$. Therefore, the overall computational complexity of Algorithm 2 is $\mathcal O({\rm max}\{n^3, K\hat{n}^6\})$ where $K$ is the number of iterations. Note that the value of $K$ is also small because the policy improvement is quadratically convergent \cite{kleinman1968iterative}. Thus, as long as $\hat{n}$ is sufficiently small, Algorithm 2 will save a lot of learning time. This will be demonstrated through numerical simulations later.

\section{Robustness Analysis}\label{sec:relax}
In the previous section, we have assumed the matrix $P$ to satisfy the condition \req{exact_P}.
In reality, however, $P$ obtained by solving \req{minpro} may not satisfy the condition exactly. 
Instead, one can only find a $P$ that satisfies this condition approximately.
Note that, since $A$ is Hurwitz, for the obtained $P$ there exists a positive number $\epsilon$ satisfying 
\begin{equation}\label{mild}
 \|(I-P^{\dg}P)x(t)\|_{\mathcal L_2} \leq \epsilon. 
\end{equation}
for any $u \in\mathcal L_2^{\rm nor}$ and a given $x_0$. 
 The following lemma is useful for analyzing the stability and the performance of the closed-loop system with a controller that is learned under this non-ideal condition.

\vspace{2mm}
 \begin{lemma}
  Consider $\Sigma$ in \req{sys}, $\hat{\Sigma}$ in \req{lowsys} and
  \begin{equation}\label{sysDelta}
   \Delta:~ \left[
\begin{array}{c}
 \dot{\hat{e}}\\
 \dot{\hat{x}}
\end{array}
\right] =
   \left[
\begin{array}{cc}
 PAP^{\dg}& PA\ol{P}^{\dg}\ol{P}\\
 0 & A
\end{array}
   \right]   \left[
\begin{array}{c}
 {\hat{e}}\\
 {\hat{x}}
\end{array}
\right] +  \left[
\begin{array}{c}
 0\\
 B
\end{array}
\right]u
  \end{equation}
  with $\hat{e}(0) = 0$, $\hat{x}(0) = x_0$. The state $x$  satisfies
 \begin{equation}\label{generalEQ}
  x = P^{\dg}\xi + \delta x,\quad \delta x := P^{\dg}\hat{e} + \ol{P}^{\dg}\ol{P}\hat{x}
 \end{equation}
  where $\xi$ obeys \req{lowsys}, for any $u(t)$ and $x_0 \in \mathbb R^n$.
 \end{lemma}
 \vspace{2mm}
\begin{proof}
 Define
  \begin{equation}
   \begin{array}{rcl}
    G(s) &\hspace{-2mm}:=&\hspace{-2mm} (sI-A)^{-1}[B~x_0]\\
    \hat{G}(s) &\hspace{-2mm}:=&\hspace{-2mm} P^{\dg}(sI-PAP^{\dg})^{-1}[PB~Px_0].
   \end{array}
  \end{equation}
 The state-space model of $\Delta(s) := G(s) - \hat{G}(s)$ is described as
 \begin{equation}\label{errorsys}
  \left\{
   \begin{array}{ccl}
\left[
\begin{array}{c}
 \dot{\hat{\xi}}\\
 \dot{\hat{x}}
\end{array}
\right] &\hspace{-2mm}=&\hspace{-2mm}
   \left[
\begin{array}{cc}
 PAP^{\dg}& 0\\
 0 & A
\end{array}
   \right]   \left[
\begin{array}{c}
 {\hat{\xi}}\\
 {\hat{x}}
\end{array}
\right] +  \left[
\begin{array}{cc}
 PB & Px_0\\
 B & x_0
\end{array}
\right] \left[
\begin{array}{c}
 u\\ \delta
\end{array}
\right]\\
\delta x&\hspace{-2mm}=&\hspace{-2mm}-P^{\dg}\hat{\xi} + \hat{x}.
   \end{array}
\right.
 \end{equation}
 Let $\hat{e} := P\hat{x} - \hat{\xi}$.
 Applying the coordinate transformation from $[\hat{\xi}^{\sf T}, \hat{x}^{\sf T}]^{\sf T}$ to  $[\hat{e}^{\sf T}, \hat{x}^{\sf T}]^{\sf T}$, the system \req{errorsys} is equivalently described as \req{sysDelta} with $\delta x$ in \req{generalEQ}.
 Since $G(s) = \Delta(s) + \hat{G}(s)$, the claim follows. \hspace{\fill} $\square$
\end{proof}
\vspace{2mm}

\begin{figure}[t]
  \begin{center}
    \includegraphics[clip,width=8.5cm]{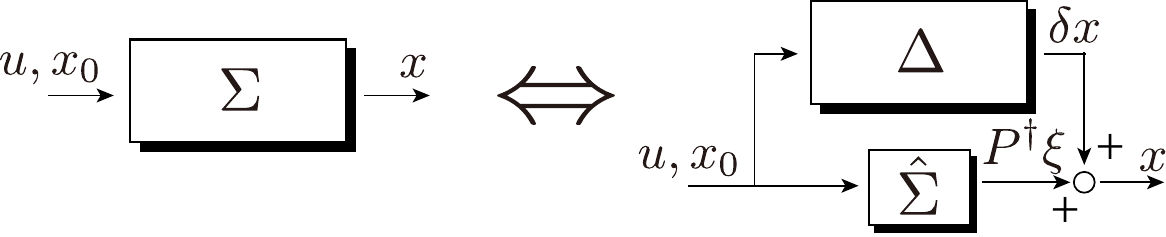}
    \caption{Relation between $\Sigma$ and the interconnection of $\hat{\Sigma}$ and $\Delta$}
   \label{blkdiagram}
   \vspace{-6mm}
  \end{center}
\end{figure}
 
 This lemma provides a tractable realization of $\Sigma$ in the form of a parallel interconnection of the low-dimensional system $\hat{\Sigma}$ and the system $\Delta$ associated with the approximation error; see Fig.~\ref{blkdiagram} for the signal-flow diagram of the interconnection. This parallel interconnection structure enables us to analyze the stability and performance of the closed-loop system using robust control theory. This is summarized as in the following theorem.

 \begin{theorem}\label{theorem3}
  Consider $\Sigma$ in \req{sys} and Algorithm 2. Let $P$  be such that $PAP^{\dg}$ is Hurwitz. Let $\hat{F}_k$ be the control gain at the $k$-th step of Algorithm 2. If
  \begin{equation}\label{sgcond}
   \epsilon < \|\hat{\Sigma}_{\rm cl}(s)\Xi(s)\|^{-1}_{\mathcal H_{\infty}}
 \end{equation}
 where $\epsilon$ follows from \req{mild}, and
 \begin{eqnarray}
  \hat{\Sigma}_{\rm cl}(s) &\hspace{-2mm}:=&\hspace{-2mm} -\hat{F}_{k}(sI - \hat{A}_{\rm F})^{-1}PB\hat{F}_{k} - \hat{F}_{k} \label{defsigma_cl} \\
  \Xi(s) &\hspace{-2mm}:=&\hspace{-2mm} P^{\dg}(sI - PAP^{\dg})^{-1}PA
 \end{eqnarray}
 with $\hat{A}_{\rm F} := PAP^{\dg}-PB\hat{F}_{k}$,
 then, $A-BF_{k}$ is Hurwitz. Furthermore, there exists $\gamma \geq 0$ such that $J$ in \req{Jx0} satisfies
 \begin{equation}\label{degra}
  J^{\frac{1}{2}} \leq \hat{J}^{\frac{1}{2}} + \gamma \epsilon 
 \end{equation}
 where $\hat{J}$ is defined in \req{Jhat}. When $\epsilon = 0$, $J = \hat{J}$. 
 \end{theorem}
\vspace{2mm}
\begin{proof}
We first prove the stability of the closed-loop system. Note that $Px = \xi + \hat{e}$ where $\hat{e}$ is defined in \req{sysDelta}. Thus, the interconnection of $\hat{\Sigma}$ with $u = -\hat{F}_{k}Px$ can be written as
 \begin{equation}
 \hat{\Sigma}_{\rm cl}:\left\{
 \begin{array}{ccl}
 \dot{\xi} &\hspace{-2mm}=&\hspace{-2mm} \hat{A}_{\rm F}\xi - PB\hat{F}_{k}\hat{e} \\
  u &\hspace{-2mm}=&\hspace{-2mm} -\hat{F}_{k}\xi - \hat{F}_{k}\hat{e}. \\
 \end{array}
 \right.
 \end{equation}
 The transfer function of $\Delta(s)$ in \req{sysDelta} from $u$ to $\hat{e}$ is
 \begin{equation}\label{utoe}
  \Delta_{ue}(s) := P^{\dg}(sI - PAP^{\dg})^{-1}PA\ol{P}^{\dg}\ol{P}(sI-A)^{-1}B.
 \end{equation}
 Note
 \begin{eqnarray}
  \|\hat{\Sigma}_{\rm cl} \Delta_{ue}\|_{\mathcal H_{\infty}} &\hspace{-2mm}=&\hspace{-2mm} \|\hat{\Sigma}_{\rm cl} \Xi \ol{P}^{\dg}\ol{P}(sI-A)^{-1}B\|_{\mathcal H_{\infty}}\nonumber \\
  &\hspace{-2mm}\leq &\hspace{-2mm}  \epsilon \|\hat{\Sigma}_{\rm cl} \Xi \|_{\mathcal H_{\infty}} < 1. \label{proof3_01}
 \end{eqnarray}
 Thus, it follows from small-gain theorem that the closed-loop of $\hat{\Sigma}_{\rm cl}$ and $\Delta_{ue}$ is stable. 

Next, we analyze the performance achieved by the control $u = -F_{k}x$. Define $Q_{\frac{1}{2}}$ and $R_{\frac{1}{2}}$ as the respective Cholesky factors of $Q$ and $R$, i.e., $Q_{\frac{1}{2}}^{\sf T}Q_{\frac{1}{2}} = Q$ and $R_{\frac{1}{2}}^{\sf T}R_{\frac{1}{2}} = R$. Define
 \begin{equation}
  y := \left[
\begin{array}{c}
 Q_{\frac{1}{2}}x\\
 R_{\frac{1}{2}}u\\
\end{array}
\right].
 \end{equation}
 The closed-loop system can then be described as the interconnection of
\begin{equation}\label{newCL}
  \left\{
\begin{array}{ccl}
 \dot{\xi}&\hspace{-2mm}=&\hspace{-2mm}\hat{A}_{\rm F}\xi + B_{\zeta}\zeta + Px_0 \delta \\
 w &\hspace{-2mm}=&\hspace{-2mm}C_{w}\xi + D_{\xi \zeta} \zeta + D_{\xi \delta}\delta \\
 y &\hspace{-2mm}=&\hspace{-2mm}C_{y}\xi + D_{y \zeta} \zeta
\end{array}
  \right.
\end{equation}
 and $\Delta$ in \req{sysDelta},  where
 $\zeta := [\hat{e}^{\sf T}, (\ol{P}^{\dg}\ol{P}\hat{x})^{\sf T}]^{\sf T}$,
 $w := [u^{\sf T}, \delta]^{\sf T}$, and
\begin{equation}\label{sys_CL}
 \begin{array}{ll}
  B_{\zeta} := \left[ -PB\hat{F}_{k}~ 0\right], &
   C_{w} = \left[-F_{k}^{\sf T} ~0\right]^{\sf T}\\
D_{\xi \zeta} :=
   \left[
    \begin{array}{cc}
     -F_{k}& 0\\ 0 & 0
    \end{array}
   \right], &
 D_{\xi \delta} := \left[
    \begin{array}{c}
     0 \\ 1
    \end{array}
   \right]\\
  C_{y} = \left[
\begin{array}{c}
 Q_{\frac{1}{2}}P^{\dg}  \\  -R_{\frac{1}{2}}\hat{F}_{k}
\end{array}
\right], &
D_{y \zeta} :=
   \left[
    \begin{array}{cc}
     Q_{\frac{1}{2}}\ol{P}^{\dg}\ol{P} &  Q_{\frac{1}{2}}P^{\dg}\\
     0 & -R_{\frac{1}{2}}\hat{F}_{k}\\
    \end{array}
   \right].
 \end{array}
 \end{equation}
 Thus
 \begin{equation}
  J^{\frac{1}{2}} = \|y\|_{\mathcal L_2} \leq \|G_{\delta y} \|_{\mathcal H_2} +
   \|G_{\zeta y} (I - \Delta G_{\zeta w})^{-1} \Delta G_{\delta w} \|_{\mathcal H_2}  
 \end{equation}
 where $G_{\circ \bullet}$ denotes the transfer function of \req{sys_CL} from the input $\circ$ to the output $\bullet$, and $\Delta$ denotes that of the system \req{sysDelta} with the output $\zeta$ in \req{newCL}.
 Note that $ \|G_{\delta y} \|_{\mathcal H_2} = \hat{J}^{\frac{1}{2}}$ and
 \begin{eqnarray}
  \hspace{-4mm}\Delta G_{\delta w} &\hspace{-2mm}=&\hspace{-2mm} \left[
\begin{array}{c}
 (sI - PAP^{\dg})^{-1}PA \\ I
\end{array}
\right] \times \nonumber \\
  && \hspace{-15mm} \ol{P}^{\dg}\ol{P}(sI -A)^{-1}\left(-B\hat{F}_{k}(sI-\hat{A}_{\rm F})^{-1}Px_0 + x_0\right).
 \end{eqnarray}
From the sub-multiplicative property of the $\mathcal H_{\infty}$-norm, we get
\begin{eqnarray}
 && \hspace{-4mm}J \leq \hat{J}^{\frac{1}{2}} + \|\mathcal G(s)\|_{\mathcal H_{\infty}} \nonumber \\
 && \hspace{-2mm}\times  \|\ol{P}^{\dg}\ol{P}(sI -A)^{-1}\left(-B\hat{F}_{k}(sI-\hat{A}_{\rm F})^{-1}Px_0 + x_0\right)\|_{\mathcal H_2}\nonumber \\
 &&\label{Jeva}
\end{eqnarray}
 where
 \[
  \mathcal G(s) := G_{\zeta y} (I - \Delta G_{\zeta w})^{-1} \Delta\left[
\begin{array}{c}
 (sI - PAP^{\dg})^{-1}PA \\ I
\end{array}
\right].
 \]
 Note that
 \begin{eqnarray}
  &&\hspace{-2mm}\|\ol{P}^{\dg}\ol{P}(sI -A)^{-1}\left(-B\hat{F}_{k}(sI-\hat{A}_{\rm F})^{-1}Px_0 + x_0\right)\|_{\mathcal H_2} \nonumber \\
  && \hspace{2mm} \leq\|\ol{P}^{\dg}\ol{P}(sI -A)^{-1}B\|_{\mathcal H_{\infty}}
   \|\hat{F}_{k}(sI-\hat{A}_{\rm F})^{-1}Px_0\|_{\mathcal H_2} \nonumber  \\
  && \hspace{2mm} + \|\ol{P}^{\dg}\ol{P}(sI -A)^{-1}x_0\|_{\mathcal H_{2}}. \label{Jeva2}
 \end{eqnarray}
 Now, we show that \req{mild} yields 
 \[
  \|\ol{P}^{\dg}\ol{P}(sI -A)^{-1}B\|_{\mathcal H_{\infty}} \leq 2\epsilon, \quad
  \|\ol{P}^{\dg}\ol{P}(sI -A)^{-1}x_0\|_{\mathcal H_{2}} \leq \epsilon.
 \]
 Let $x^{(x)}(t) := e^{At}x_0$ and $x^{(u)}(t) :=  \int_0^{t} e^{A(t-\tau)}Bu(\tau) d\tau.$ Note that \req{mild} holds for $u \equiv 0$. Thus, $\|\ol{P}^{\dg}\ol{P}x^{(x)}\|_{\mathcal L_{2}} \leq \epsilon$, which implies $  \|\ol{P}^{\dg}\ol{P}(sI -A)^{-1}x_0\|_{\mathcal H_{2}} \leq \epsilon$. Since $x^{(u)} = x - x^{(x)}$, it follows that $\|\ol{P}^{\dg}\ol{P}x^{(u)}\|_{\mathcal L_2} \leq \|\ol{P}^{\dg}\ol{P}x\|_{\mathcal L_2} + \|\ol{P}^{\dg}\ol{P}x^{(x)}\|_{\mathcal L_2} \leq 2\epsilon$ for any $u \in \mathcal L_2^{\rm nor}$. This is equivalent to $\|\ol{P}^{\dg}\ol{P}(sI -A)^{-1}B\|_{\mathcal H_{\infty}} \leq 2\epsilon$. 
 From \req{Jeva} and \req{Jeva2}, \req{degra} follows where 
 \begin{equation}
  \gamma := \|\mathcal G(s)\|_{\mathcal H_{\infty}}(1 + 2\|\hat{F}_{k}(sI-\hat{A}_{\rm F})^{-1}Px_0\|_{\mathcal H_2}).
 \end{equation}
 This completes the proof.  \hspace{\fill} $\square$
\end{proof}


We briefly discuss a set of situations where \req{sgcond} holds and $PAP^{\dg}$ is Hurwitz. In general, $PAP^{\dg}$ tends to be stable as $\hat{n}$ gets larger because $A$ is Hurwitz. Moreover, it is known that $PAP^{\dg}$ is always stable for $A=A^{\sf T}$ (e.g., bidirectional networks composed of first-order subsystems) \cite{ishizaki2013model} or for $A$ as Metzler (e.g., positive directed networks) \cite{ishizaki2015clustered}. Note here that $P^{\dg} = P^{\sf T}$ holds if $P$ is chosen by left-singular vectors of $X$ in \req{defXX}. Similarly, it is clear from \req{defsigma_cl} that $\|\hat{\Sigma}_{\rm cl}\|_{\mathcal H_{\infty}}$ gets smaller as $\|\hat{F}_k\|$ is smaller. Thus, from \req{proof3_01} and the fact $\|\hat{\Sigma}_{\rm cl} \Xi \|_{\mathcal H_{\infty}} \leq \|\hat{\Sigma}_{\rm cl}\|_{\mathcal H_{\infty}} \|\Xi\|_{\mathcal H_{\infty}}$, we can see that \req{sgcond} holds for low-gain controllers, which can be achieved by choosing the input weight $R$ in \req{Jx0} to be large. Using a low-gain controller, however, may result in a poor $\hat{J}$. This can be counteracted by choosing a $P$ that makes the approximation error $\epsilon$ in \req{mild} small, thereby improving the ideal cost $\hat{J}$.

When these assumptions are satisfied, Theorem 3 implies the following two results: first, the learned controller at the $k$-th step of Policy Improvement in Algorithm 2 guarantees the stability of the closed-loop system. Second, smaller this error closer is the closed-loop performance cost to the ideal cost $\hat{J}$. 



\section{Extension to semi-stable systems}\label{sec:semi}
We next extend Algorithm 2 to semi-stable systems. For simplicity of analysis, we assume that $A$ in system \req{sys}, although unknown, has one semi-stable eigenvalue. This is often the case when the network exhibits a consensus property over a connected graph \cite{mesbahi2010graph}. The proposed approach can be easily generalized to systems with multiple semi-stable poles.
We denote the eigenvector associated with the semi-stable pole as $v$. The main assumption behind the proposed method is that $v$ must be known. This assumption is actually true for many real-world applications. For consensus networks, $v$ is the vector of all ones. For power system networks, where the semi-stable pole arises from Kirchoff's current balance, $v$ contains identical entries corresponding to the generator phase angles while all other entries are zero. This is because the dynamics of the generators are coupled through the difference of their phase angles, and therefore the eigenvector $v$ that changes all the angles uniformly does not have any influence on the dynamics. Moreover in many of these networks the control objective is driven by the goal of controlling relative states to quantify how one agent is performing with respect to another. In such situations, a common assumption that is satisfied is:
 \begin{equation}\label{kercon}
  \ker Q \supseteq \im~v. 
 \end{equation}
 Starting from these observations, we state the following proposition.

\vspace{2mm}
 \begin{proposition}
Consider a semi-stable system $\Sigma$ in \req{sys}, whose semi-stable eigenvector is denoted by $v$.
  Construct $P_{\rm c} \in \mathbb R^{\hat{n} \times (n - 1)}$ such that
 \begin{equation}\label{newPcon}
  (I - P_{\rm c}^{\dg}P_{\rm c})\ol{v}^{\dg}x(t) \equiv 0, \quad \forall t
 \end{equation}
  for any $u$ and a fixed $x_0$. Define
 \begin{equation}\label{defPc}
  P = P_{\rm c}\ol{v}^{\dg}.
 \end{equation}
  Suppose that \req{rankcond2} is satisfied.  If $Q$ in \req{Jx0} satisfies \req{kercon}, 
then by applying Algorithm 2 to the semi-stable system $\Sigma$ the following two statements hold:
 \begin{enumerate}
  \item[i)] $A-BF_k$ is semi-stable whose semi-stable eigenvector is $v$.
  \item[ii)] The control $u=-F_{\infty}x$ minimizes $J$ in \req{Jx0}.
 \end{enumerate}
 \end{proposition}
\vspace{2mm}
  \begin{proof}
   Define $\eta := {v}^{\dg}x$ and $\ol{\eta} := \ol{v}^{\dg}x$. Applying a coordinate transformation from $x$ to $[\ol{\eta}^{\sf T}, \eta^{\sf T}]^{\sf T}$, we have
   \begin{equation}\label{defeta}
    \left[
     \begin{array}{c}
      \dot{\ol{\eta}} \\ \dot{\eta}
     \end{array}
    \right] =
    \left[
    \begin{array}{cc}
     \ol{v}^{\dg}A\ol{v} & 0\\
     {v}^{\dg}A\ol{v} & 0
    \end{array}
    \right]     \left[
     \begin{array}{c}
      {\ol{\eta}} \\ {\eta}
     \end{array}
    \right] +     \left[
     \begin{array}{c}
      \ol{v}^{\dg}B \\ v^{\dg}B
     \end{array}
    \right]u,
   \end{equation}
   where the relation $Av = 0$ is used. Note that $\ol{v}^{\dg}A\ol{v}$ is Hurwitz. Similar to Lemma 1, it follows that
   \[
   {\rm im} P^{\sf T} = {\rm im}~\mathcal R(\ol{v}^{\dg}A\ol{v}, ~\ol{v}^{\dg}[B~ x_0]).
   \]
   Therefore, applying coordinate transformation from $\ol{\eta}$ to $[\xi^{\sf T}, \ol{\xi}^{\sf T}]^{\sf T}$ where $\ol{\xi} := \ol{P}x$, we get \req{lowsys}. The compressed state $\xi$ obeys \req{lowsys}. From the upper-triangular structure of \req{lowsys1} and the fact that \req{lowsys1} is stable, we can see that $PAP^{\dg}$ is Hurwitz.
   Furthermore, we denote the Cholesky factor of $Q$ by $Q_{\frac{1}{2}}$, i.e., $Q = Q_{\frac{1}{2}}^{\sf T}Q_{\frac{1}{2}}$.    Since \req{kercon} yields $Q_{\frac{1}{2}}v = 0$,  we have
   \[
    Q_{\frac{1}{2}}x = Q_{\frac{1}{2}}(\ol{v}\ol{\eta} + v\eta) = Q_{\frac{1}{2}}\ol{v}(P_{\rm c}^{\dg}\xi + \ol{P}_{\rm c}^{\dg}\ol{\xi}) = Q_{\frac{1}{2}}P^{\dg}\xi,
   \]
   where the relation $\ol{\xi}(t) \equiv 0$ is used. Thus, $x^{\sf T}Qx = \xi^{\sf T}\hat{Q}\xi$.
 Similar to the proof of Theorem 1, it is proven that
 $PAP^{\dg} - PB\hat{F}_k$ is Hurwitz for any $k$.
 Note that the eigenvalues of $A-BF_k$ with $F_k$ in \req{defK} are identical to those of $PAP^{\dg}-PB\hat{F}_k$, $\ol{P}A\ol{P}^{\dg}$, plus a zero eigenvalue. Thus, $A-BF_k$ is semi-stable. Also note that
 \[
  (A-BF_k)v = -B\hat{F}_kP_{\rm c}\ol{v}^{\dg}v = 0.
 \]
   Thus, $v$ is the eigenvector corresponding to the semi-stable pole of $A-BF_k$. Therefore, claim i) follows.
   Claim ii) follows by using the same argument as in the proof of Theorem 1. This completes the proof. \hspace{\fill}$\square$
  \end{proof}
  \vspace{2mm}
  
This proposition implies that as long as $P_{\rm c}$ satisfies \req{newPcon}, i.e. the data compression is lossless, Algorithm 2 with $P$ defined in \req{defPc} can provide an optimal controller for semi-stable systems. Such a matrix $P_{\rm c}$ can be found by the singular value decomposition of
\[
 X_{\rm c} = \ol{v}X
\]
where $X$ is defined in \req{defXX}. Furthermore, even for the non-ideal case when $\|(I - P_{\rm c}^{\dg}P_{\rm c})\ol{v}^{\dg}x(t)\| \leq \epsilon$, Theorem 3 with $P$ defined in \req{defPc} still holds for semi-stable systems. 

\section{Numerical Simulation}\label{sec:sim}
\subsection{Case 1: Consensus Network System}\label{subsec_sim01}

\begin{figure}[t]
  \begin{center}
    \includegraphics[clip,width=8.8cm]{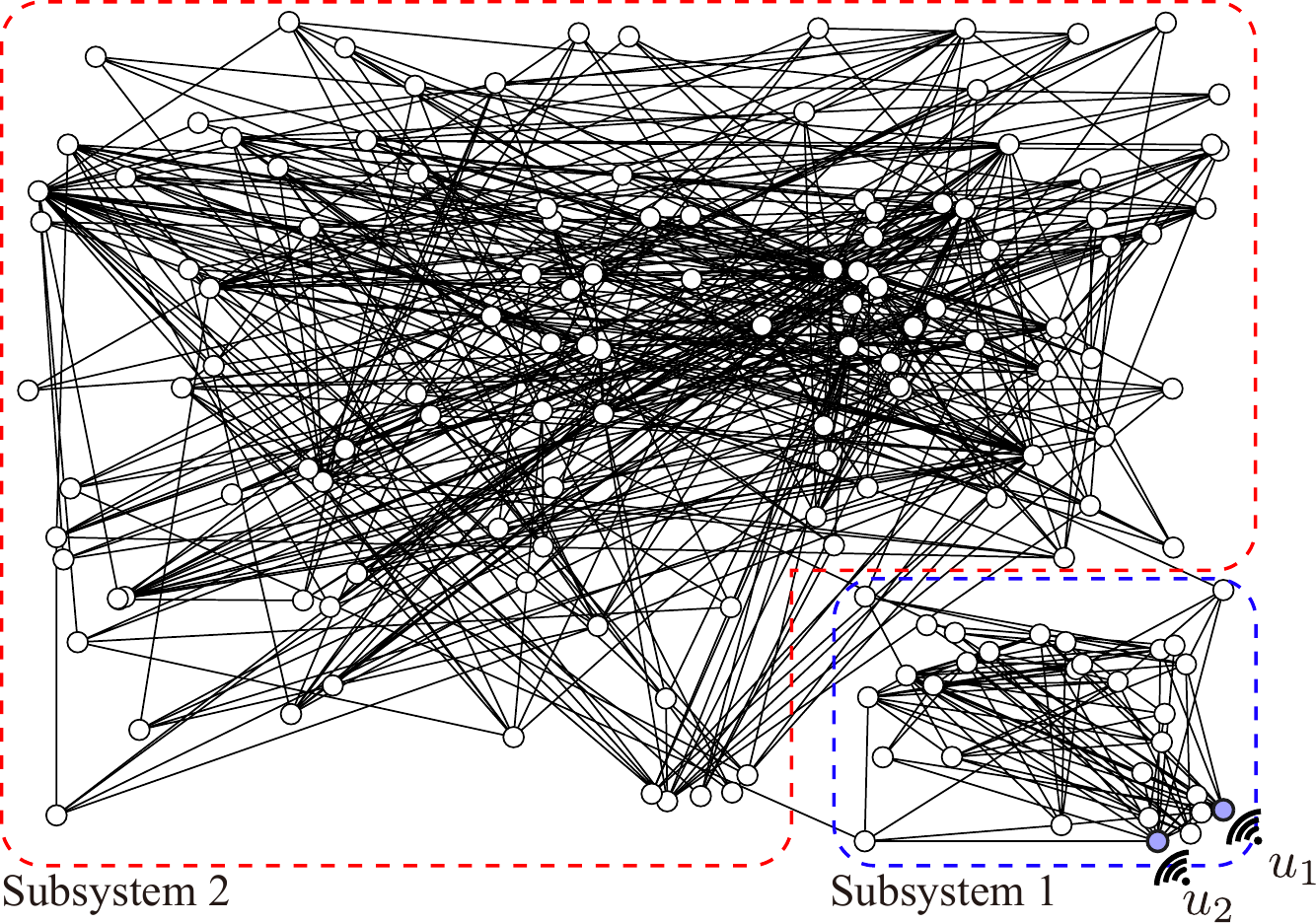}
    \caption{A consensus network system composed of 150 nodes}
   \label{con2}
  \end{center}
 \vspace{-4mm}
\end{figure}

\begin{figure}[t]
  \begin{center}
    \includegraphics[clip,width=8.8cm]{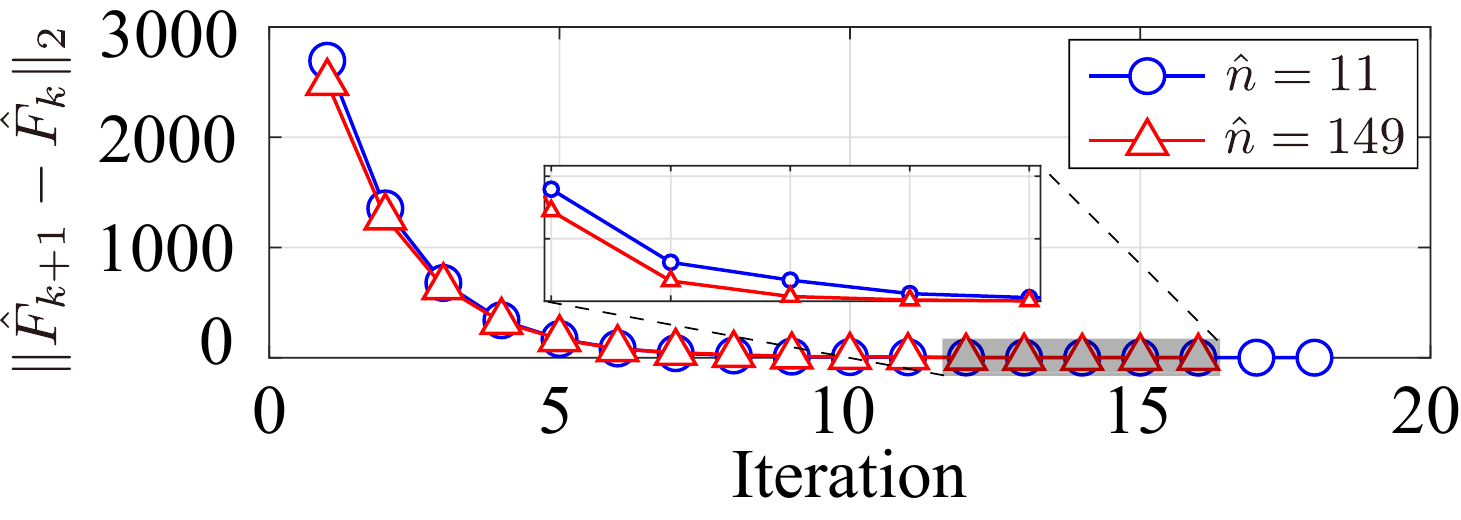}
    \caption{The variation of the change of $\hat{F}$ }
   \label{Kit}
  \end{center}
       \vspace{-4mm}
\end{figure}

\begin{figure}[t]
  \begin{center}
    \includegraphics[clip,width=8.8cm]{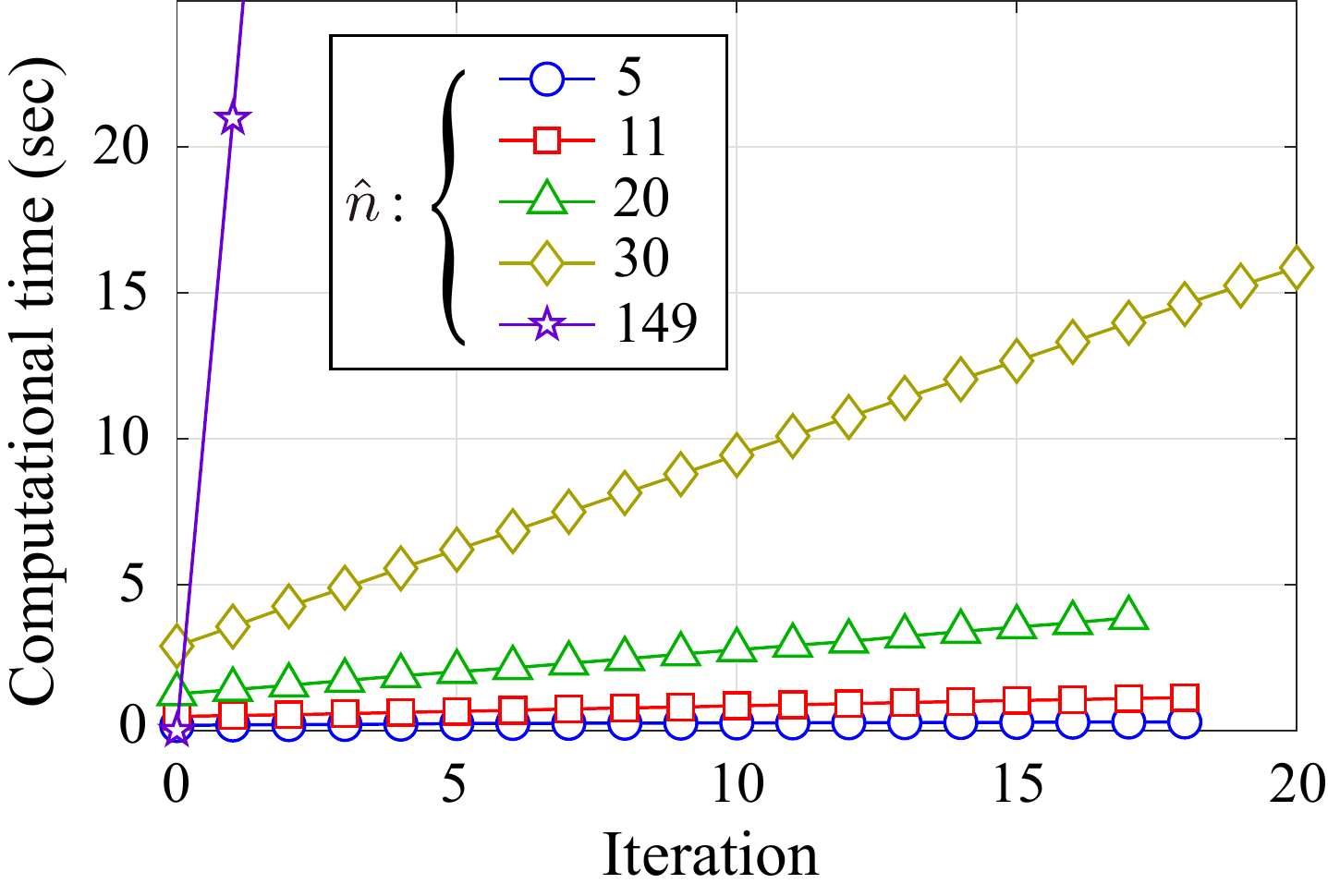}
    \caption{Computational time needed for controller design}
    \label{CT}
  \end{center}
 \vspace{-4mm}
\end{figure}

\begin{figure}[t]
  \begin{center}
    \includegraphics[clip,width=8.8cm]{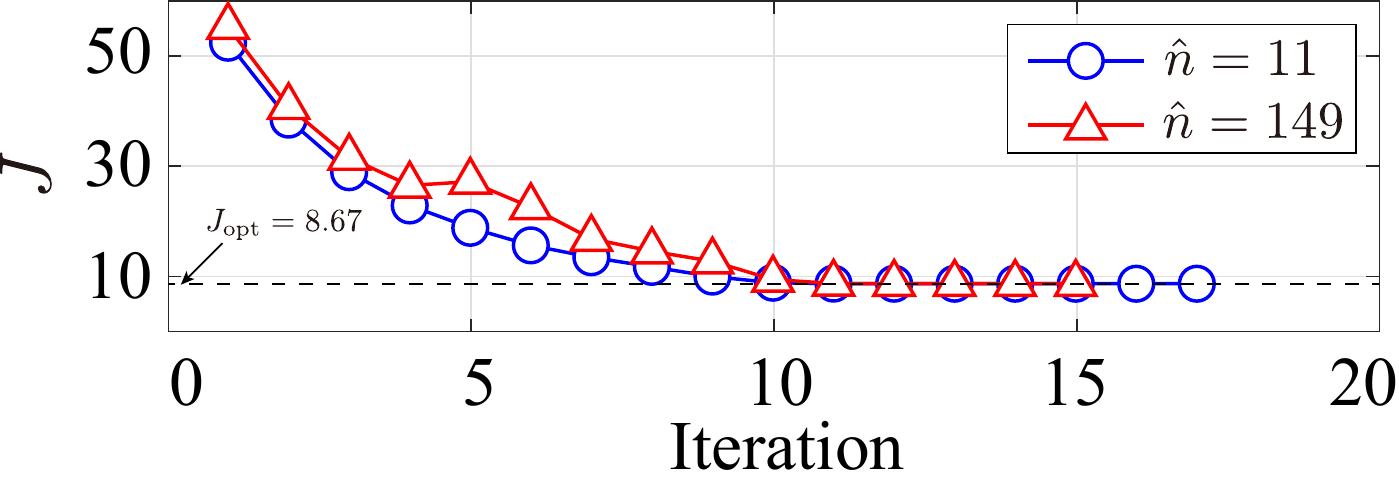}
    \caption{The variation of the performance as the increase of iteration, where $J_{\rm opt}$ is the cost achieved by the model-based optimal controller. }
   \label{J_it}
  \end{center}
       \vspace{-4mm}
\end{figure}

\begin{figure}[t]
  \begin{center}
    \includegraphics[clip,width=8.8cm]{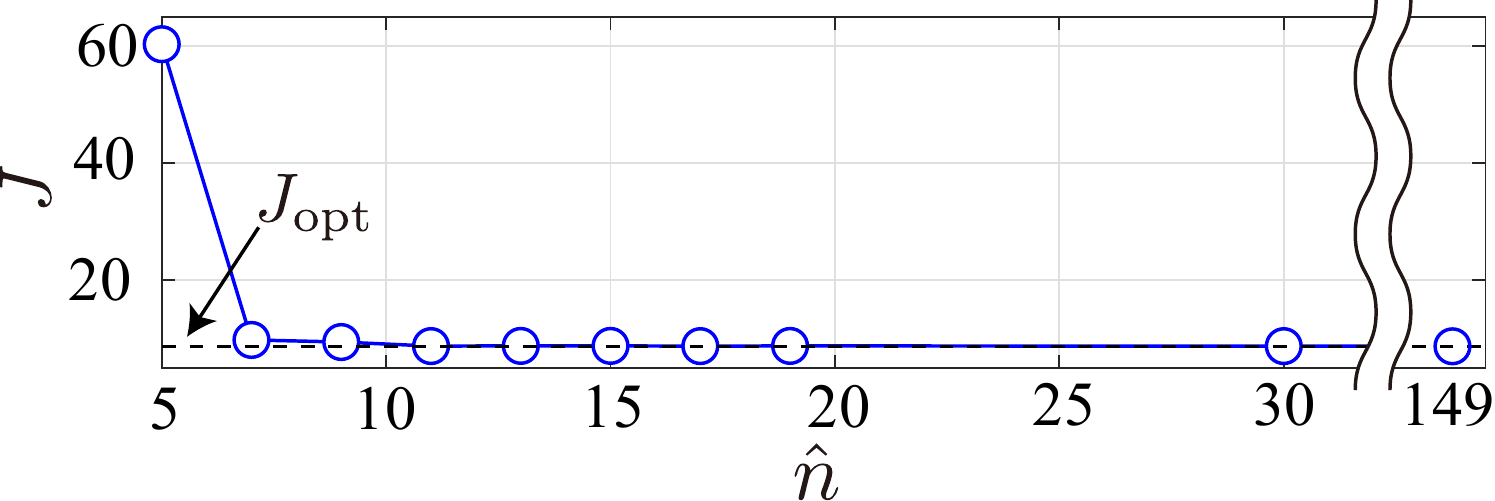}
    \caption{Cost versus the dimensionality}
    \label{Jcurve}
  \end{center}
       \vspace{-4mm}
\end{figure}

  \begin{center}
 \begin{table}[t]
  \begin{center}
  \begin{tabular}{|c|c|c|} \hline
    \multicolumn{1}{|c|}{w/o control}
      & \multicolumn{1}{c|}{With control in case where $\hat{n}=11$} & \multicolumn{1}{c|}{With $F_{\rm opt}$}\\ \hline
   0.0000&   0.0000 &   0.0000 \\  
   -0.0127&   -0.2631&   -0.2615 \\
   -0.2735&  -0.2991&   -0.2787 \\
   -0.2958&   -0.3770&   -0.3396 \\ 
   -0.3769&   -0.4022&   -0.3769 \\  \hline
  \end{tabular}
   \label{table2}
   \vspace{1mm}
   \caption{First five dominant eigenvalues of the open-loop system and closed-loop system}
  \end{center}
  \vspace{-4mm}
 \end{table}
  \end{center}

  \begin{figure}[t]
  \begin{center}
    \includegraphics[clip,width=8.8cm]{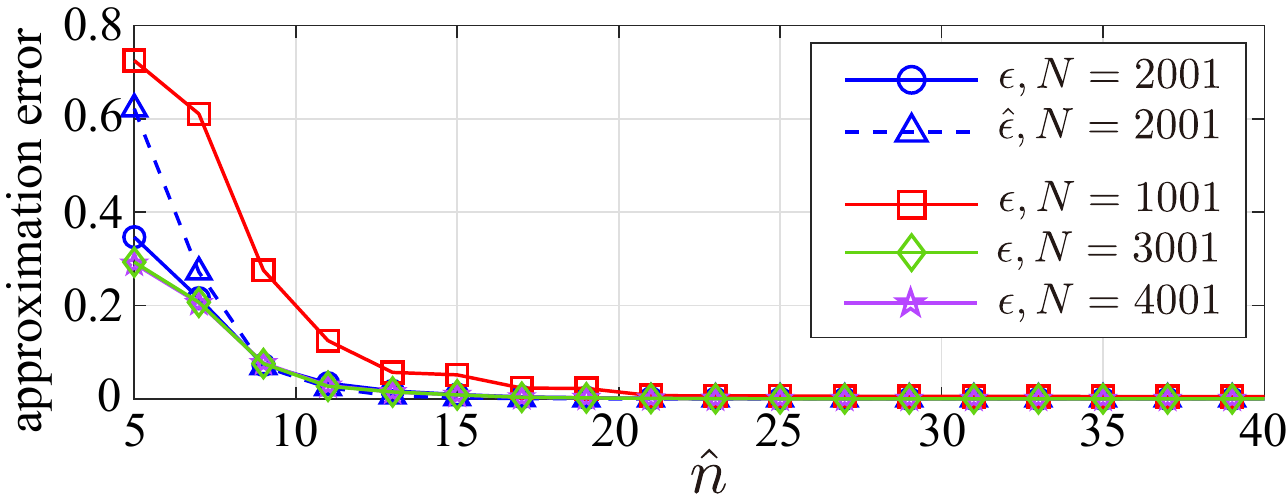}
    \caption{The approximation error $\epsilon$ in \req{def_ehat} versus $\hat{n}$}
    \label{DimCheck}
  \end{center}
       \vspace{-4mm}
\end{figure}

 We investigate the effectiveness of the proposed algorithm through an example of  a consensus network as shown in Fig.~\ref{con2}.
 The network is composed of two subsystems (areas) consisting of $30$ and $120$ nodes, respectively. The fact that the network consists of two areas is only used for describing the network structure and its dynamics. We will not utilize this fact for controller design.  The dynamics of the $i$-th node is described as 
\begin{equation}\label{consensus}
 \dot{x}_i = \sum_{j \in \mathbb N_i}a_{i,j}(x_j - x_i) + b_i u_i, 
\end{equation}
where $x_i \in \mathbb R$ is the state, and $\mathbb N_i$ is the index set of nodes connected to the $i$-th node. 
The inter-area graph structure is given as a Barabasi-Albert model \cite{albert2002statistical}, and the weights of the edges are randomly chosen from the range $(0, 0.5]$.
The two areas are connected through four links between four nodes in each area. The link weights are all assigned to be $0.1$. Let control inputs be applied only to the first two nodes of Area 1, i.e., $b_1=b_2 = 1$ and $b_i = 0$  for $i \in \{3, \ldots, 150\}$. Clearly, the system \req{consensus} can be described as \req{sys} with $n=150$. Note that this network system is semi-stable, and the corresponding eigenvector is ${\bf 1}_n$. 

The control objective here is to improve the speed of consensus among all nodes without knowing $a_{ij}$ and $b_i$ for all $\{i,j\}$. To this end, we take $Q$ in \req{Jx0} such that $x^{\sf T}Qx = \alpha \sum_{i=1}^{n} (x_{1} - x_{i})^2$ with a scalar weight $\alpha$, and $R = I$. Let $\alpha = 50$. 
The sampling sequence $\{t_j\}$ in the {\it Online Data Collection} stage is taken as $t_j = 0.01j$ for $j \in \{0,\ldots, 2000\}$. From $t=0$ to $t=1$ (sec), we inject the noise $u(t) = \beta \sum_{k=1}^{400}\sin(w_k t)$ with $w_k$ randomly chosen from the range $[-20,20]$ for $k\in \{1,\ldots, 400\}$. Let $\beta = 0.05$. 

We first find $P$ as given in \req{defPc}, by using the collected data $\{x(t_j)\}_{j \in \{0, \ldots, 2000\}}$. Since the consensus network system is semi-stable, we project the data onto the stable subspace. Let $\overline{v}^{\dg} \in \mathbb R^{(n-1) \times n}$ be such that $\overline{v}^{\dg} {\bf 1}_n = 0$. We compute $\overline{\eta} = \overline{v}^{\dg}x$. Next, similarly to \req{minpro}, for a given $\hat{n}$ we solve 
$ \min_{P_{\rm c} \in \mathbb R^{\hat{n} \times (n-1)}} \sum_{j=0}^{N-1} \|(I-P_{\rm c}^{\dg}P_{\rm c}) \ol{\eta}^{\dg}(t_j)\|^2.$  
Let $\hat{n} = 11$. We conclude the preconditioning step in Algorithm 2 by computing $\hat{\phi}$, $\hat{\rho}$, and $\hat{\sigma}$ in \req{phihat}-\req{sigmahat}. The state-feedback gain $\hat{F}_k$ is computed by running the Policy Improvement step.
Fig.~\ref{Kit} shows the variation of $\hat{F}_k$ in \req{sol2} for $\hat{n} = 11$ and 149.
Under the criterion that the algorithm is stopped when $\|\hat{F}_{k+1} - \hat{F}_k\| \leq 0.01$, 
it takes 16 iterations to converge when $\hat{n} = 11$, and 18 when $\hat{n} = 149$. In other words, the convergence speed for $\hat{n} = 11$ is almost the same as that for $\hat{n} = 149$. 
Furthermore, the resultant control performance for $\hat{n}=11$ is $J = 8.70$ while that achieved by the optimal model-based controller is $J_{\rm opt} = 8.67$. This result shows that the obtained controller for $\hat{n} = 11$ achieves nearly optimal control performance. 

Next, we investigate the computational complexity of Algorithm 2. All computations were done using MATLAB 2017b, in an Intel(R) Core(TM) i7-4587U 3.00GHz, RAM 16.0GB computer. Fig.~\ref{CT} shows the computational time taken for the Policy Improvement portion of Algorithm 2, for different values of $\hat{n}$. The computational time at the initial iteration is the time used for preconditioning. The purple line with stars corresponds to the case where Algorithm 1 without data compression is applied, i.e., $P_{\rm c} = I$. The line shows that as high as 21 seconds of real-time are needed for each step of Policy Improvement. On the other hand, the blue, red, green, and yellow lines with circles, squares, triangles, and diamonds show the case for different values of $\hat{n} \leq n$. By comparing those with the purple line, we can see that the dimensionality reduction efficiently reduces the computational time at every step. For example, the total computation time without data compression is 337 seconds whereas that with reduction of $\hat{n}=20$ is only 4 seconds.
Fig.~\ref{J_it} shows the convergence of the performance objective $J$ for $\hat{n} = 11$ and 149. In both cases at least 10 iterations are needed for achieving nearly optimal control performance. On the other hand, the computational time for $\hat{n} = 149$ is 175 seconds while that for $\hat{n}=11$ is only 0.6 seconds. These results confirm that the data compression drastically reduces the learning time.

Fig.~\ref{Jcurve} shows the cost $J$ for different choice of controllers learned for different values of $\hat{n}$. From this figure, we can see that there exists a trade-off relation between the dimensionality (i.e., computational time) and closed-loop performance. For comparison, we design an optimal controller gain $F_{\rm opt}$ by solving a Riccati equation for the original system model \req{sys}. 
In Fig.~\ref{Jcurve}, the dashed line shows the cost $J_{\rm opt}$ achieved by this ideal controller. We can see that the controller with $\hat{n}=11$ is almost optimal. Furthermore, Table~\ref{table2} shows the first five dominant eigenvalues of the open-loop system and the closed-loop system when $F_{18} = \hat{F}_{18}P$ and $F_{\rm opt}$ are used, respectively. Clearly, the second dominant eigenvalue, associated with the consensus speed, is drastically improved by the proposed RL controller.

We close this subsection by investigating how the matrix $P$ obtained by solving \req{minpro} approximates the ideal relation \req{exact_P}. 
One choice of $\epsilon$ satisfying \req{mild} is given as
\begin{equation} \label{def_ehat}
 \scalebox{1.0}{$\displaystyle
  \epsilon := \|\ol{P}^{\dg}\ol{P}(sI-A)^{-1}B\|_{\mathcal H_{\infty}} + \|\ol{P}^{\dg}\ol{P}(sI-A)^{-1}x_0\|_{\mathcal H_{2}}
  $}
\end{equation}
where $P$ is constructed from \req{minpro}. Clearly, the condition \req{exact_P} is equivalent to \req{mild} with $\epsilon = 0$. Thus, the parameter $\epsilon$ in \req{def_ehat} can be a measure that can quantify how accurately $P$ in \req{minpro} approximates \req{exact_P}. Considering that $\{x(t_j)\}_{j \in \{0, \ldots, 2000\}}$ (i.e., $N = 2001$) is used for formulating \req{minpro}, we compute $\epsilon$ for different choices of $\hat{n}$. In Fig.~\ref{DimCheck}, the blue line with circles show the resultant $\epsilon$ for $\hat{n}$. We can see that the value of $\epsilon$ in \req{def_ehat} becomes smaller as $\hat{n}$ increases. Moreover, when $\hat{n} = 29$, the resultant $\epsilon = 1.5 \times 10^{-4}$, which implies that the obtained $P$ in this case is almost ideal.
This tendency is also observed when the endpoint of the sampling index $j$ is changed to 1000, or 3000, or 4000. Furthermore, we can see that the variation of $\epsilon$ over $\hat{n}$ saturates after a certain point as $N$ increases. This is because the original controllable subspace $\mathcal R(A, [B,x_0])$ is fixed and the controllable subspace captured by the data becomes closer to the original as $N$ increases. 
The exact value of $\epsilon$ is, however, difficult to compute using only the acquired data $\{x_j\}$ because $A$ and $B$ are unknown. One alternative metric instead of $\epsilon$ in \req{def_ehat} can be
\begin{equation}\label{def_epshat}
 \hat{\epsilon} := \sqrt{{\rm tr} ((\ol{P}^{\dg})^{\sf T}X\ol{P})}
\end{equation}
where $X$ is defined in \req{defXX}. The reason why this can be an alternative is that when the input $u$ is white Gaussian and $x_0 = 0$, as $N$ increases we can expect that the value $\hat{\epsilon}$ converges to $\|\ol{P}^{\dg}\ol{P}x\|_{\mathcal L_2}$. In Fig.~\ref{DimCheck}, we show $\hat{\epsilon}$ by the red line with triangles. By comparing this and the blue line, we can see that indeed $\hat{\epsilon}$ in \req{def_epshat} can be a guideline for choosing $\hat{n}$. 



\vspace{3mm}
\subsection{Case 2: Power System Model}\label{subsec_sim02}
\begin{figure}[t]
  \begin{center}
    \includegraphics[clip,width=8.8cm]{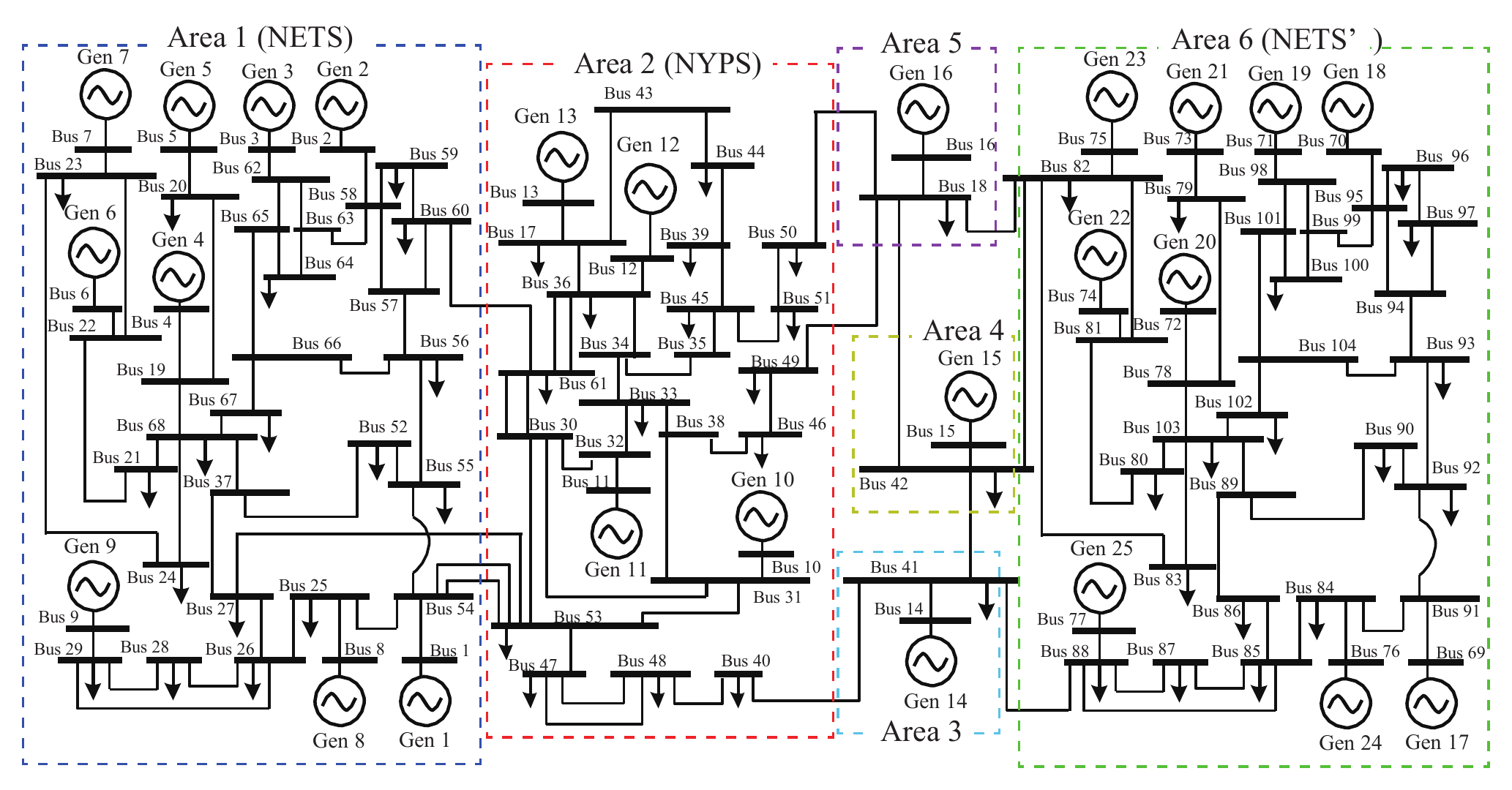}
    \caption{Power system model consisting of 25 generators, 52 loads, and 104 buses. } 
    \label{power} 
  \end{center}
       \vspace{-4mm}
\end{figure}
\begin{figure*}[t]
  \begin{center}
    \includegraphics[clip,width=18cm]{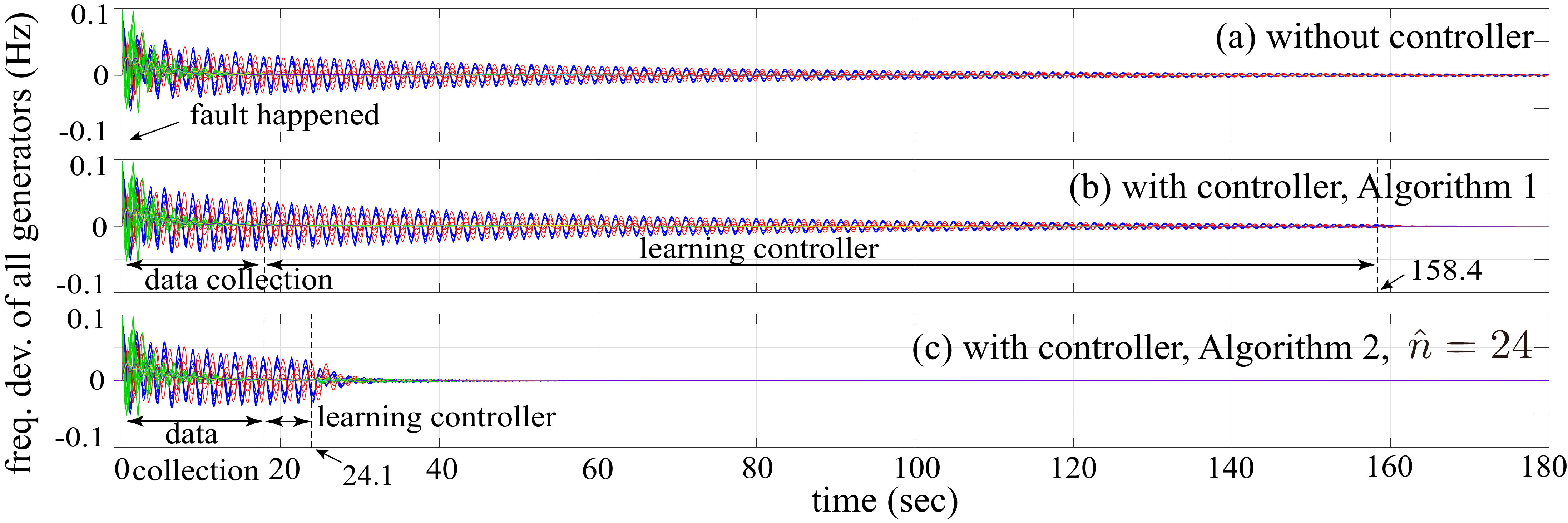}
    \caption{Transient response of frequency deviation of generators (a) without control, (b) by the optimal controller designed by Algorithm 1, and (c) by the proposed controller in Algorithm 2. } 
    \label{68busresult} 
  \end{center}
       \vspace{-4mm}
\end{figure*}

Next, we investigate how the proposed reinforcement learning controller can be effectively applied for wide-area oscillation damping of an electric power system without the knowledge of its small-signal model. 
We consider a power system model composed of six areas, as shown in Fig.~\ref{power}. Note that these areas need not be coherent. 
The part containing areas 1 through 5 is the standard IEEE 68-bus model \cite{pal2006robust}, while Area 6 is a repetition of Area 1. This area models the New England test system (NETS). The overall model consists of 25 synchronous generators and 52 loads, interconnected via 104 buses. The loads are modeled as constant impedance loads. Following \cite{csm}, the dynamics of the power system can be modeled as a nonlinear differential-algebraic equation 
\begin{equation}\label{model68bus}
 \dot{x} = f(x, V, w), \quad 0 = g(x, V)
\end{equation}
where $x := [x_1^{\sf T}, \ldots, x_{25}^{\sf T}]^{\sf T} \in \mathbb R^{100}$, $x_{k} \in \mathbb R^{4}$ is the state of the $k$-th generator (angle, frequency deviation, internal voltage, and field voltage), $V := [V_1, \ldots, V_{104}]^{\sf T} \in \mathbb C^{104}$, $V_{k}$ is the complex-valued voltage of the $k$-th bus, $w := [w_1, \ldots, w_{25}]^{\sf T} \in \mathbb R^{25}$ and $w_k \in \mathbb R$ is the control input which enters the automatic voltage regulator (AVR) as an additional reference voltage signal to the $k$-th generator. We suppose that the state $x$ is available from state estimators based on Phasor Measurement Units (for details please see \cite{singh2014decentralized}).
Due to the double integrator in the swing equations, the small-signal model corresponding to \req{model68bus} is semi-stable, with the eigenvector for the zero eigenvalue being $v = {\bf 1}_{25} \otimes e_1^{4}$.
We consider a fault, which is modeled as an impulse input reflected through the change in the initial conditions of the states from their pre-disturbance equilibrium. The frequency deviations of generators in Areas 1-6 after this fault are depicted by the blue, red, cyan, yellow, purple, and green lines in Fig.~\ref{68busresult}(a), respectively. We can see from this subfigure that the inter-area oscillations (whose frequency lies in the range 0.1-2.0 Hz) appear dominantly in the time-responses of the generator frequencies. To enhance the damping performance of this power system, we consider designing a model-free reinforcement learning controller. Although the controller is designed based on a linearized model, it is implemented using the original nonlinear model \req{model68bus}.

We consider that the control input of every generator inside the $l$-th area, where $l \in \{1,\ldots, 6\}$, is equal. That is, $w_{i} = u_{[l]}$ for any $i$ in the index set of the generators belonging to the $l$-th area. Let $u = [u_{[1]}, \ldots, u_{[6]}]^{\sf T}$. The control input is determined by
\begin{equation}
 u = F(x - x^{\star})
\end{equation}
where $x^{\star}$ is the setpoint of $x$ in \req{model68bus}. Note that this setpoint is a priori known by measuring the steady state without exploiting any detailed knowledge about the system model. The controller gain $F$ is determined from Algorithm 2 replacing $x$ with $x - x^{\star}$. The design parameters are chosen as follows. $Q$ in \req{Jx0} is chosen such that $x^{\sf T}Qx = \alpha \sum_{l=1}^6 \omega_{[l]}^2$ where $\omega_{[l]}$ is the average frequency deviation of generators inside the $l$-th area and $\alpha$ is a scalar weight. We choose $\alpha=10$ and $R = I$ in \req{Jx0}.
From $t=0$ to $t=2$ (sec), we inject the exploration noise described in \ref{subsec_sim01} to each generator through its excitation system voltage $w_k$. The noise amplitude is taken as $\beta = 10^{-3}$. The data is collected for time $t \in [0, 18]$ with a sampling period of $0.01$ seconds.
The total time for executing Algorithm 1 is found to be $140.4$ seconds. The designed controller is implemented in the system starting from time $t = 18+140.4 = 158.4$ (sec). Fig.~\ref{68busresult}(b) shows the trajectories of the frequency deviations of all generators after the implementation of the controller. By comparing Figs.~\ref{68busresult}(a) and (b), one can see that the damping performance over the entire time-horizon of interest is very limited due to the long learning time (140.4 seconds). To reduce the computation time, we apply the Algorithm 2 to the system with $\hat{n} = 24$. The learning time in that case reduces to only 6.1 seconds. Fig~\ref{68busresult}(c) shows the case where the designed controller is implemented at time $t = 24.1$ (sec). It can be clearly seen that the performance improves drastically compared to Figs.~\ref{68busresult}(a)-(b). Note that the controller is linear while the power system model used for the simulations is described by the nonlinear model \req{model68bus}. This demonstrates the inherent robustness property of the proposed RL controller for this example.


 \section{Conclusion}\label{sec:con}
    In this paper, we proposed a new RL control strategy for large-scale network systems. The proposed control strategy can avoid the curse of dimensionality by utilizing the low-rank property of the network in terms of its controllabilty. In the proposed strategy, (almost) uncontrollable state variables are eliminated by projecting the state data on to the controllable subspace.
    This projection is achieved in a completely model-free way using only measured data. Results are validated by two practical examples, both of which show notable speed-up of the learning time while guaranteeing satisfactory sub-optimal performance.
   The main contribution of the paper is to show how two individually well-known concepts in dynamical system theory and machine learning - namely, model reduction and reinforcement learning - can be combined together to construct a real-time control design that can be highly useful for extreme-scale networks. We have shown a trade-off relation between the improvement of learning time and the cost of degradation of the closed-loop performance due to the data compression error, answering the two questions that were raised in the introduction.

  Our future work will include the extension of this method to a hierarchical structure where microscopic or local controllers and macroscopic or global controllers are learned in parallel using the data compression concept. We also wish to extend this approach to a stochastic setting using discrete-time Markov decision processes (MDPs), and to nonlinear input-affine systems as the recent paper \cite{kashima2016noise} shows that the SVD of the matrix $X$ in \req{defXX} can capture the controllable subspace of such nonlinear systems.
 

 \section*{Acknowledgment}
 This research was partially supported by CREST, JST Grant Number JPMJCR15K1, Japan.
 The work of the first author was partly supported by JSPS Grant-in-Aid for Research Activity Start-up JP19K2350. 

 

\ifCLASSOPTIONcaptionsoff
  \newpage
\fi
\bibliographystyle{IEEEtran}
\bibliography{ref}

\end{document}